\documentclass[journal]{new-aiaa} 
\usepackage[utf8]{inputenc}
\usepackage{todonotes}
\usepackage{graphicx}
\usepackage{amsmath}
\usepackage{subfigure}
\usepackage{longtable,tabularx}
\usepackage{lineno}
\usepackage{svg}
\usepackage{natbib}
\title{Predictions of flow distortions inside a \\ serpentine diffuser from large-eddy simulations}

\author{Rahul Agrawal \footnote{Currently: Lead Software Engineer, Cadence Design Systems, San Jose, CA 95134, USA. AIAA Member. } }
\affil{Center for Turbulence Research, Stanford University, Stanford, CA 94305, USA}

\author{Chad Winkler \footnote{Boeing Technical Fellow. Associate Fellow AIAA. } }
\affil{The Boeing Company, St. Louis, MO 63110, USA}

\author{Sanjeeb T. Bose\footnote{Distinguished Engineer, Cadence Design Systems. Adjunct Professor, ICME Stanford. Associate Fellow AIAA. }}
\affil{Cadence Design Systems, San Jose, CA 95134, USA}
\affil{Institute for Computational and Mathematical Sciences, Stanford University, Stanford, CA 94305, USA}

\author{Parviz Moin \footnote{Franklin P. and Caroline M. Johnson Professor, Dept. of Mechanical Engineering, Stanford University. Fellow AIAA. } }
\affil{Center for Turbulence Research, Stanford University, Stanford, CA 94305, USA}

\begin{document}
\maketitle

\begin{abstract}

This work examines the flow separation and the resulting pressure distortions at the exit plane of a serpentine diffuser operating at both subsonic and transonic conditions. Wall-modeled large-eddy simulations (WMLES) using the charLES flow solver are performed at three exit-plane Mach numbers, $Ma_{AIP} \approx \{ 0.36, \, 0.46, \, 0.54 \}$. First, it is shown that the onset of flow separation inside a serpentine diffuser may likely experience strong, non-local history effects. A grid refinement study consisting of five grids (from 30 million to 3 billion cells) is conducted for all Mach numbers. 
The recently proposed dynamic tensor-coefficient Smagorinsky subgrid-scale \citep{agrawal2022non} and sensor-aided non-equilibrium wall models \citep{agrawal2024non} compare 
favorably with experimental measurements for the pressure recovery and azimuthal flow distortion at all Mach numbers.  
The pressure recovery and azimuthal 
flow distortion are predicted to within $0.3\%$ and $7\%$, respectively, which are both within the experimental error bounds. However, the simulations underpredict the maximum azimuthal distortion in comparison to the experiments for all conditions. Statistical comparisons of the dynamic azimuthal flow distortions suggest that the present LES reasonably captures the ring-averaged mean distortions, and the statistical distributions of the distortion around the mean.  Extreme 
events are underestimated by the present simulations, potentially highlighting that significantly longer integration times may be necessary. \\



\end{abstract}

\section{Nomenclature}

{\renewcommand\arraystretch{1.0}
\noindent\begin{longtable*}{@{}l @{\quad=\quad} l@{}}
$\rho$ & density \\
$e$ & resolved internal energy \\
$E$ & sum of resolved internal and kinetic energies \\
$T$ & temperature \\
$p$ & pressure \\
$u_i$ & velocity components \\
$S^d_{ij}$ & deviatoric component of the strain-rate tensor \\
$\tau^{sgs}_{ij}$ & subgrid-scale stress tensor \\
$Q^{sgs}_{j}$ & subgrid heat-flux \\
$l_p$ & pressure-gradient imposed viscous length scale \\
$\Delta$ & size of computational grid \\
$\Delta^+$ & resolution based on friction velocity \\
$\Delta_w^+$ & wall-resolution based on friction velocity \\
$\Delta_p^+$ & wall-resolution based on pressure-gradient imposed length scale \\
AIP  & Aerodynamic Interface Plane \\
$D$ & diameter of Aerodynamic Interface Plane \\ 
$W_{40}$ & mass flow rate through the AIP \\
$Ma_{AIP} $ &  Mach number at the AIP \\
$Ma_{t} $ &  Mach number at the throat \\
$L_x$ & streamwise length of the domain \\ 
$U_{\infty}$ & freestream velocity \\
$u_{\tau}$ & friction velocity \\
$u_p$ & pressure-gradient based velocity \\
$Pr$ and $Pr_t$  & Molecular and turbulent Prandtl numbers respectively \\
$Re_\theta$ & Reference Reynolds number based on momentum thickness \\
$Mcv$ & Million Control Volumes \\
$G$ & Green's function \\
$C_p$ & mean surface pressure coefficient \\
$p^0_t$ & reference inlet total pressure  \\
$\gamma $ & specific heat ratio  \\
$\chi $ & time-averaged activity of wall-model sensor \\
$\mathrm{PR}$ & temporally and face-averaged pressure recovery at the AIP \\
$\mathrm{dPcP}_i$ & ring-averaged azimuthal distortion at the AIP \\
$\mathrm{dPcP}$ & face-averaged azimuthal distortion at the AIP \\
WMLES & Wall-modeled Large-eddy simulation \\
LES & Large-eddy simulation \\
RANS & Reynolds-averaged Navier-Stokes \\
DDES & Delayed detached-eddy simulation \\
Q-Q & quantile-quantile plots \\
EVT & Extreme value theory \\

\end{longtable*}}

\section{Introduction }
 
\noindent
The flow inside a serpentine diffuser is relevant for the design of inlet systems in high-speed propulsion applications, for instance, in embedded engines or engines with low length-to-diameter ratios \citep{burrows2020evolution}. Such inlet geometries have varying cross sectional areas with curvature gradients within the diffuser duct, which can lead to boundary-layer separation. In these circumstances, a reduced total pressure recovery and increased flow distortion at the engine face result, which may limit the engine's operability and performance.  \\

\noindent
Typically, the effect of the flow distortion within diffusers is measured at the aerodynamic interface plane (AIP), placed close to the diffuser's exit, where the uniformity of the recovered flow is calculated. Generally, as the Mach number at the throat of the diffuser increases (due to a higher incoming mass flow), the 
pressure recovery can be reduced due to the onset of flow separation. Previously, several propulsion aerodynamics workshops \citep{delot2013comparison,winkler2017summary} have focused on simulating a serpentine diffuser to assess whether 
these flow phenomena could be accurately predicted. These workshops have shown a significant spread in the prediction of the flow recovery and distortion metrics from Reynolds-averaged Navier-Stokes (RANS) simulations and delayed detached eddy simulations (DDES). In particular, \citet{lakebrink2018numerical} showed that at transonic throat Mach numbers, both RANS and DDES underpredict flow separation on the diffuser walls.  \\

\begin{figure}
    \centering
    (a) \includegraphics[width=0.4\textwidth]{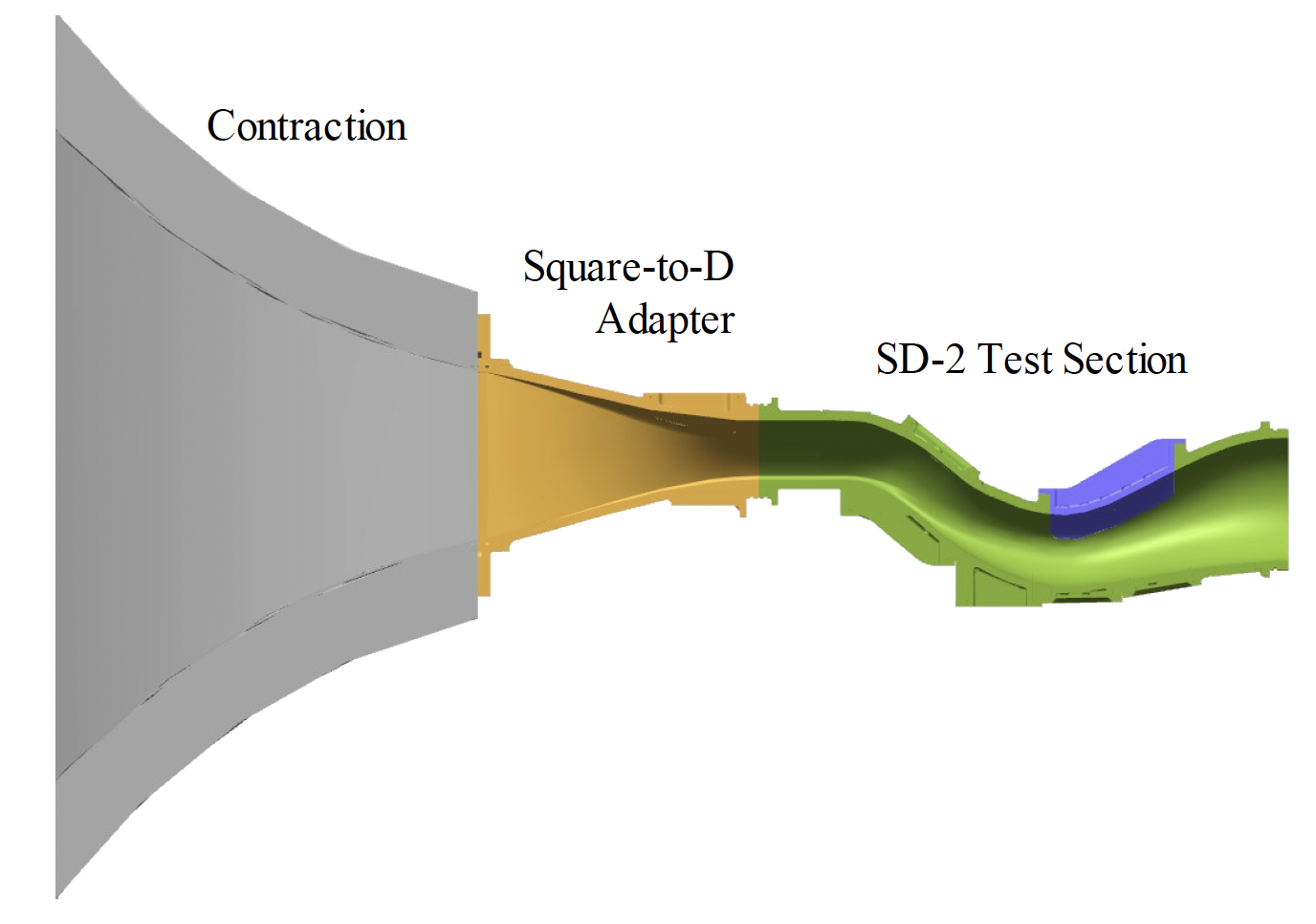}  
    (b)   \includegraphics[width=0.4\textwidth]{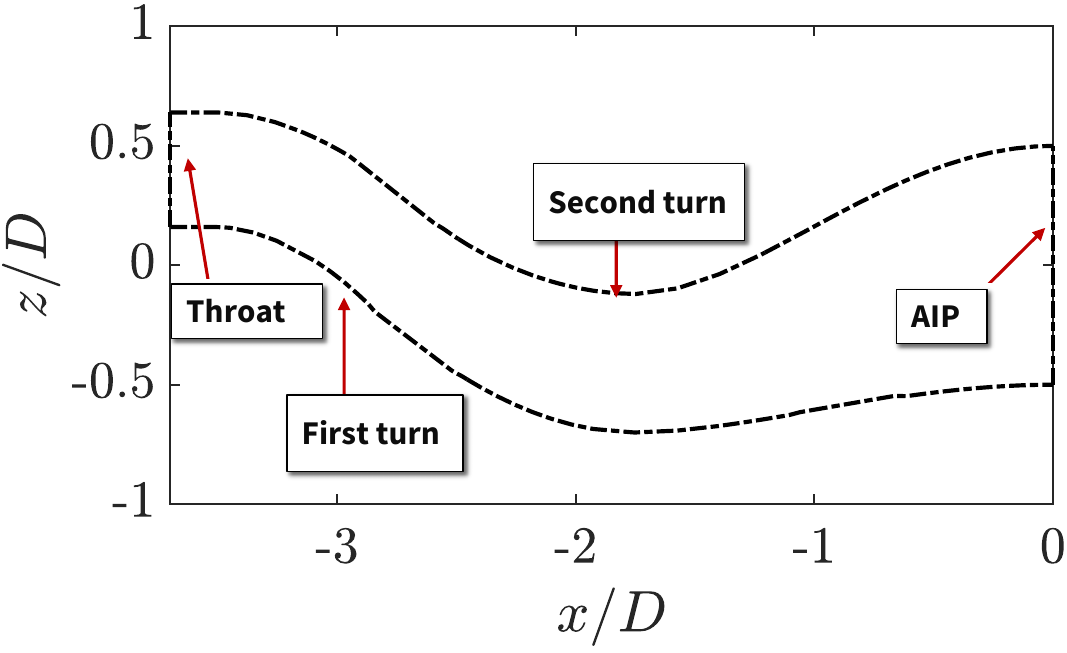} 
    \vspace{10pt} \\
    (c) \includegraphics[width=0.4\textwidth]{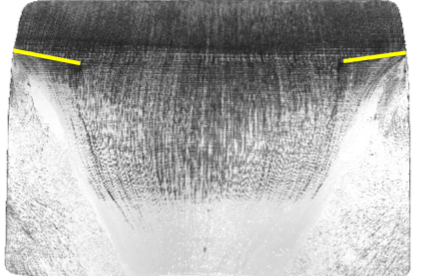}
    \caption{ (a) A sectional view of the SD-2 test section attached to the upstream contraction and the square- to D-shaped adapter [image reproduced from \citet{burrows2020evolution}]. (b) A geometric schematic of the test section of the SD-2 diffuser ($D$ is the diameter of the duct at the exit plane) and (c) the surface oil visualization [reproduced from \citet{burrows2020evolution}] of the SD-2 diffuser at $Ma_{AIP} \sim 0.54$, where the black dotted lines mark the horseshoe separated regions, and the yellow solid lines mark the shock wave footprints.}
    \label{fig:sd2geom}
\end{figure}

\noindent
In this work, we simulate the transonic diffuser, SD-2, studied experimentally by \citet{burrows2020evolution} at Georgia Tech. This choice is motivated by the recent RANS, DDES, and wall-modeled LES (WMLES) studies of \citet{lakebrink2018numerical}, \citet{lakebrink2019toward} and \citet{stahl2023modal}, which have all reported significant challenges in the prediction of mean flow recovery at the diffuser exit. Figure \ref{fig:sd2geom}(a) shows a schematic of the experimental geometry with the 106:1 contraction region that accelerates the incoming flow into the test section. Figure \ref{fig:sd2geom}(b) shows a simplified schematic of only the test section. \citet{lakebrink2019toward} and \citet{stahl2023modal} have previously predicted the static pressure inside the duct but reported over-predictions of the recovery at the exit plane. Further, \citet{lakebrink2018numerical} showed that RANS (using a Spalart-Allmaras model) strongly overpredicts the mean recovery due to underpredicted flow separation. This under-prediction was concentrated on the lower wall of the diffuser, where the combined effect of the shock-induced and smooth-wall separation dictates the resulting AIP pattern. For DDES, \citet{lakebrink2018numerical} reported deteriorating predictions of the mean flow recovery with increasing mass flow. \\

\noindent
In light of these previous investigations, we examine the predictive performance of wall-modeled LES in capturing the mean flow recovery, distortions at three flow rates for the SD-2 case. In particular, we focus on the transonic ($Ma_t \approx 0.7$ at the throat) case, where previous RANS, DDES and wall-modeled LES have mispredicted the flow characteristics at the AIP. At these conditions, the incoming flow at the throat is approximately at an $Re_\theta \approx 6000$. Figure \ref{fig:sd2geom}(c) shows the shock footprints that form on the shoulders of the bottom wall at the first turn; the shock front moves toward the centerline as it moves downstream. The detached shear layer then reattaches downstream at $x/D \approx -1.5$. At the second turn, the boundary layer separates on the top wall due to the adverse pressure gradient at $x/D \approx -1.5$ before reattaching at $x/D \approx -1 $.  \\

\noindent
The rest of this article is written as follows: Section \ref{sec:charles} describes the numerical framework used
in this investigation. Section \ref{sec:greenfunction} provides an \emph{a-priori} assessment of the non-local effects encountered in this flow via Green's function-based approach. The simulation setup and boundary conditions are discussed in Section \ref{sec:setup}. Section \ref{sec:ma0p70wmles} provides the \emph{a posteriori} comparisons of the predicted mean surface pressures, flow recovery and distortion at the AIP for the transonic flow (throat Mach number, $Ma_t \approx 0.7$). In Section \ref{sec:machsens}, we present the effect of Mach number on surface pressure, flow recovery and distortions. We also compare the predicted dynamic distortions with those from the experiments for all Mach number flows considered. Finally, concluding remarks are provided in Section \ref{sec:conclude}. 

\section{Numerical framework}
\label{sec:charles}

\noindent
The simulations presented herein are performed using charLES\footnote{from Cadence Design Systems, San Jose, CA}, an explicit, unstructured, finite-volume solver for the compressible Navier-Stokes equations. This code is formally second-order accurate in space and third-order accurate in time and utilizes Voronoi-diagram-based grids. Formally, skew-symmetric operators are employed to conserve kinetic energy in the inviscid, zero Mach number limit 
and the scheme also approximately preserves entropy in the inviscid, adiabatic limit. More details of the solver and relevant validation studies can be found in \citet{bres2018large}, \citet{goctransoniccrm}, \citet{agrawal2023reynolds} and \citet{agrawal2025modeling}. \\  
A brief description of the previously mentioned compressible Navier-Stokes equations for LES is provided as follows. If the filtered, large-scale fields (such as velocity and pressure) in LES are denoted by $\overline{f}$, and their corresponding Favre average is denoted by $\widetilde{f}$, then the resulting equations of a compressible flow [of internal energy $e$, density $\rho$,  temperature $T$, viscosity $\mu(T)$ and thermal conductivity $\kappa(T)$] and velocity vector $\vec{u} \; = \; \{ u_1, \;u_2, \;u_3\}$ are given as, 

\begin{equation}
\frac{\partial \overline{\rho} }{\partial t }
+ \frac{\partial (\overline{\rho} \; \widetilde{u}_i )}{\partial x_i} = 0  
\end{equation}

\begin{equation}
\frac{\partial  (\overline{\rho} \; \widetilde{u}_i )}{\partial t }+\frac{\partial (\overline{\rho} \; \widetilde{u}_j \;  \widetilde{u}_i )}{\partial x_j } =- \frac{\partial \overline{p}}{\partial x_i } + \frac{\partial (\mu \widetilde{S^d}_{ij} ) }{\partial x_j } -\frac{\partial \tau^{sgs}_{ij}}{\partial x_j} ,
\end{equation}

and

\begin{equation}
\frac{\partial  \overline{E}}{\partial t }+\frac{\partial (\overline{E} \; \widetilde{u}_j)   }{\partial x_j } =- \frac{\partial (\overline{p}\; \widetilde{u_i}) }{\partial x_i } + \frac{\partial   (\mu \widetilde{S^d_{ij}} \widetilde{u_i} )}{\partial x_j } -\frac{\partial (\tau^{sgs}_{ij} \widetilde{u_i})  }{\partial x_j} - \frac{\partial Q_j^{sgs} }{\partial x_j } + \frac{ \partial }{\partial x_j } ( \kappa \frac{\partial \overline{T}}{\partial x_j }) ,
\end{equation}
where $\overline{E} =\overline{\rho}\widetilde{e} + 0.5 \;  \overline{\rho}\widetilde{u_i}\widetilde{u_i}$ is the sum of the resolved internal and kinetic energies. $\widetilde{S^d}_{ij}$ is the deviatoric part of the resolved strain-rate tensor. The relationship between the temperature and the molecular viscosity is assumed to follow a power law with an exponent of 0.75. A constant molecular Prandtl number approximation ($Pr = 0.7$) allows computation of the thermal conductivity. Two additional terms, $\tau^{sgs}_{ij}$ and $Q^{sgs}_{j}$ require modeling closure. The subgrid stress tensor, $\tau^{sgs}_{ij}$ is defined as  $\tau^{sgs}_{ij} = \overline{\rho } (\widetilde{u_i u_j} - \widetilde{u}_i  \widetilde{u}_j) $. Similarly, $Q_j^{sgs} = \overline{\rho} (\widetilde{e u_j} - \widetilde{e}  \widetilde{u_j} ) $ is the subgrid heat flux. In this work, the isotropic component of the subgrid stress is absorbed into pressure, leading to a pseudo-pressure field. The subgrid heat flux is modeled using the constant turbulent Prandtl number approximation ($Pr_t = 0.9$) applied to the dissipative component of the subgrid-stress tensor. \\

\noindent
In typical wall-modeled LES, a shear stress, and a heat flux boundary value is supplied to the  LES solver to close the discrete system of governing equations. In this work, in particular, we assess both the conventional dynamic Smagorinsky subgrid-scale closure \citep{moin1991dynamic} and equilibrium wall closures \citep{cabot2000approximate,lehmkuhl2018large} 
(hereby referred to as DSM/EQWM models, used in several external aerodynamics prediction workshops \citep{hlcrmkonrad,agrawalarb2023_2,kiris2023hlpw}) and the dynamic tensor coefficient subgrid-scale \citep{agrawal2022non} with a sensor-aided non-equilibrium wall closure \citep{agrawal2024non} (hereby referred to as DTCSM/SWM models). The latter choice of models is motivated from recent investigations \citep{agrawal2022non,agrawal2024non,agrawal2025qinetiq} that have demonstrated improved grid convergence and prediction accuracy in mildly separated flow regimes, where the flow experiences non-negligible history effects (also see Section \ref{sec:greenfunction}). For both wall models, it is remarked that a first-point matching approach is used since the current flow solver has not shown any evidence of a ``log-layer mismatch"  \citep{kawai2012wall,yang2017llm} in the simulation of turbulent channel flows in the range, $ 1000 \leq Re_{\tau} \leq 4200$ with typical wall-modeled LES resolutions. \\

\section{Flow nonlocality measure via Green's function}
\label{sec:greenfunction}

\noindent
As an \emph{a priori} measure, recently, \citet{agrawal2023reynolds} utilized a Green's function based analysis (for the pressure-Poisson system) for determining regions that influence the onset of a turbulent separation. This approach was shown to be qualitatively consistent with integral measures of flow separation \citep{agrawal2023thwaites} for the flow over the NASA/Boeing speed bump \citep{uzun2021high}. Here, we extend this analysis to an internal flow geometry to qualitatively analyze the extent of non-local flow effects. \\

\noindent
We summarize the approach here for completeness. Consider the time-averaged Poisson equation (of an incompressible boundary layer) for pressure near the separation point.
\begin{equation}
     \rho  \langle \frac{\partial \bar{u}_i }{\partial x_j} \rangle \langle \frac{\partial \bar{u}_j } {\partial x_i}  \rangle +  \rho  \langle \frac{\partial \bar{u}^{\prime} _i }{\partial x_j} \frac{\partial \bar{u}^{\prime} _j } {\partial x_i}  \rangle  =  - \nabla^2 \langle \bar{p} \rangle . 
    \label{eqn:laplace}
\end{equation}
where $\bar{\cdot}$ and $\langle \cdot \rangle$ denote the filtered 
LES fields and a time average operator, respectively, and homogeneous Neumann boundary conditions are applied on the pressure field at the solid wall. \citet{agrawal2023grid} reasoned that the second term on the left side of Equation \ref{eqn:laplace} is of the same order as the first term (resulting from the mean flow gradients) in the vicinity of a separation point (also see Figs. 18-20 in \citet{uzun2021high}). Thus, the following approximation holds,    

\begin{equation}
    2  \rho \langle \frac{\partial \bar{u}_i }{\partial x_j} \rangle \langle \frac{\partial \bar{u}_j } {\partial x_i}  \rangle   \approx  - \nabla^2 \langle \bar{p} \rangle . 
\end{equation}

\begin{figure}[!ht]
\centering
   (a) {\includegraphics[width=0.45\textwidth]{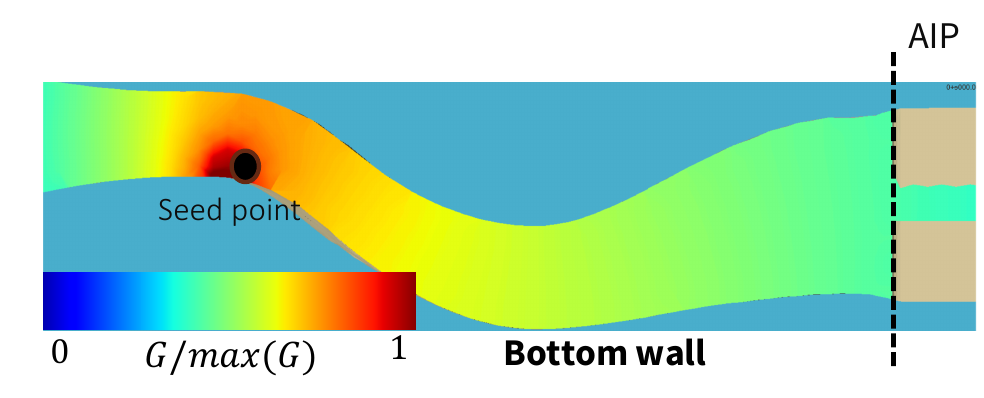}}       
   (b) {\includegraphics[width=0.45\textwidth]{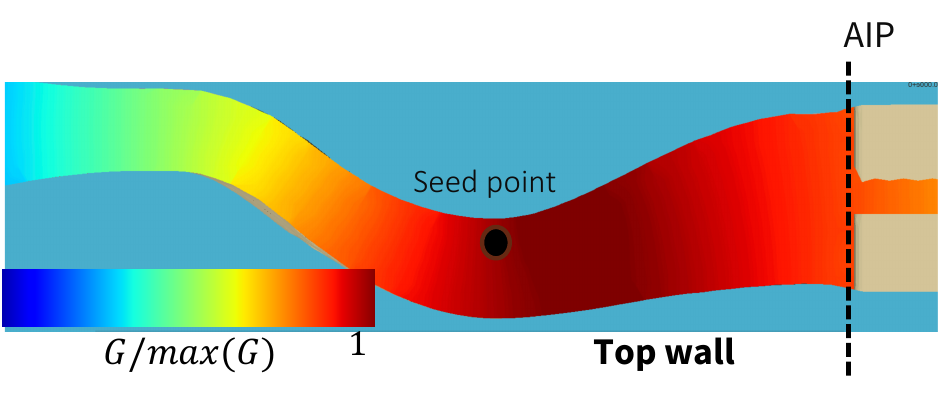}}        
    \caption{ Contours (at the midspan) of the Green's function ($G$) field for the flow inside the SD-2 diffuser with respect to the mean separation point on (a) the bottom wall and (b) the top wall.    }
    \label{fig:sd2gf}
\end{figure}

\noindent
In an \emph{a priori} sense, we utilize the Green's function of this system to highlight potential regions which may influence the pressure at the point of flow separation. Discretely, this includes the solution of Laplace equation with Neumann boundary conditions for a given flow domain; mathematically given as the solution of the following system   \\

\begin{equation}
    \frac{\partial^2 G}{\partial x_i \partial x_i} = \delta(x-x^*) \hspace{10pt} \mathrm{with \, boundary \, conditions } \hspace{10pt} \frac{dG}{dn} = 0, 
\end{equation}
where $x^*$ is a point corresponding for which the Green's function is computed and $n$ is the normal direction corresponding to a solid boundary. We choose several 
locations at the approximate separation points (one each on both walls, determined from the flattening of the experimental $C_p$ trace at the midspan). Figure \ref{fig:sd2gf} shows that the Green's function field (denoted as $G$) corresponding to the location on the bottom wall decays faster than the corresponding top wall field. On the bottom wall, the separation is likely most affected by the flow just upstream (i.e., the occurrence of the shock) and the flow turning just downstream. However, the spatial support of this field is non-compact (for instance, the field only decays to 50\% of the maximum value even at the AIP). It may be then conjectured that the separation on the bottom wall interacts with the flow on the top wall through the spreading of its wake in the core. These potentially indicate that the flow separation is affected by both upstream and downstream perturbations, and refined grid-resolutions may be required to resolve relevant flow scales in the majority of the geometry (consistent with the present gridding approach, see Section \ref{sec:setup}).  

\section{Simulation setup and boundary conditions}
\label{sec:setup}
\begin{figure}[!ht]
    \centering
    (a) \includegraphics[width=0.5\textwidth]{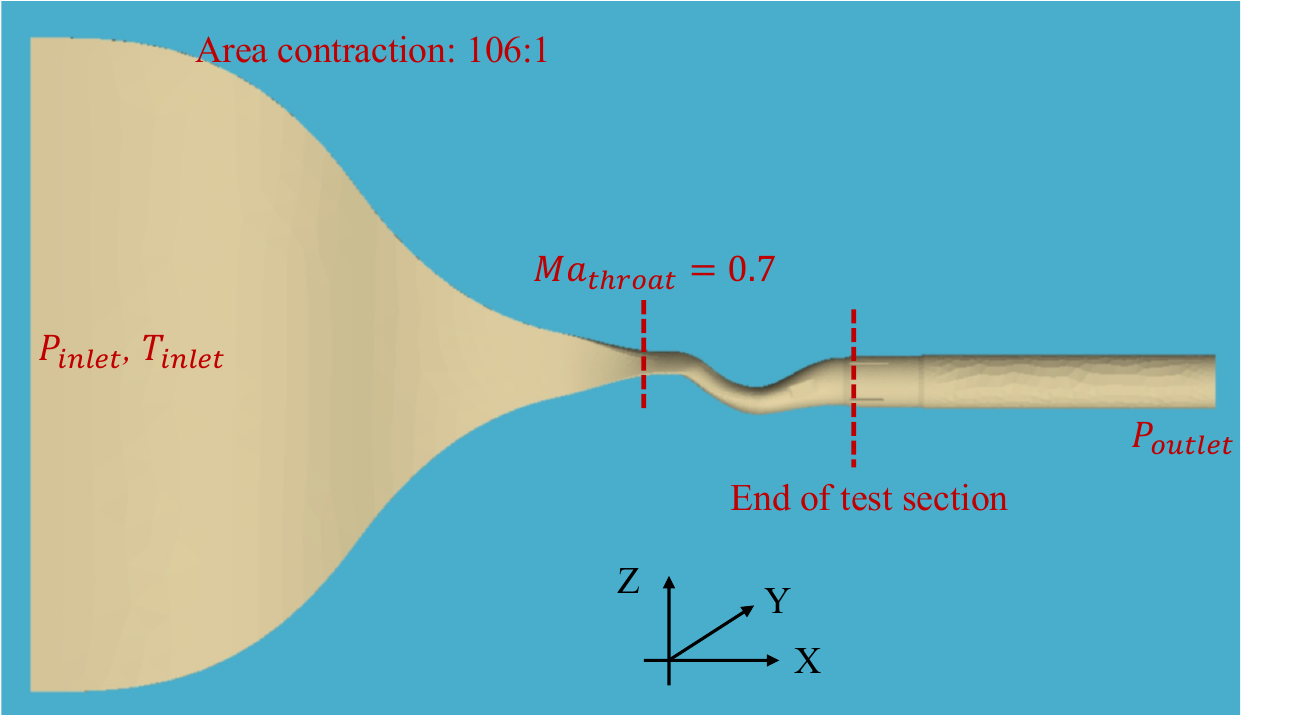} \\
    (b) \includegraphics[width=0.9\textwidth]{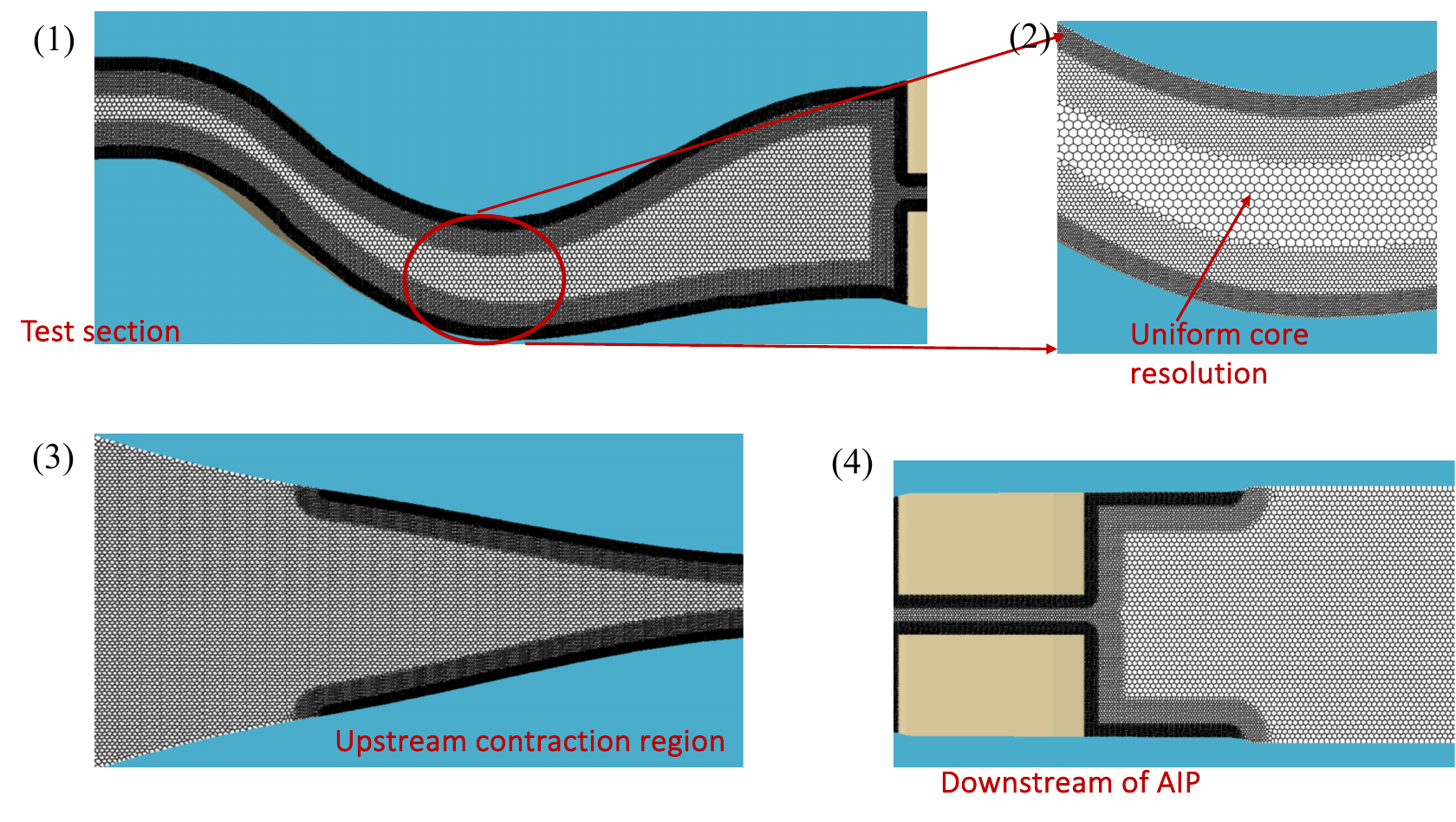}
    \caption{  (a) A schematic of simulation setup, including the bellmouth contraction of area ratio 106:1 that leads into the test section and the AIP. (b) A schematic of the Voronoi grid arrangement, corresponding to the $L3$ grid, in the test section, upstream contraction region and downstream (of AIP) regions.   }
    \label{fig:schemticsd2}
\end{figure}

\noindent
Figure \ref{fig:schemticsd2} shows a schematic of the computational setup and a two-dimensional cross section of the Voronoi gridding lattice. In line with the previous WMLES \citep{stahl2023modal} and DES studies \citep{lakebrink2019toward}, the 106:1 contracting bellmouth inlet is included, followed by the test section, and an additional cylindrical test section to serve as a flow outlet.  \\

\noindent
The walls of the diffuser are refined isotropically relative to the core resolution, in factors of two per refinement step (see Figure \ref{fig:schemticsd2}(b) for a schematic of the grids). The core region of the diffuser is uniformly meshed, with the grid size adapting with each refinement step (see Table \ref{table:sd2grids}). As a useful metric, on the finest grid, $L7$, the average inner resolution (in viscous units) on both the walls is $\Delta_w^+ \approx 15$, and in the core, $\Delta^+ \approx 120$. In terms of the length scale imposed by the strongest inviscid pressure gradient [$l_p$, see \citet{agrawal2023reynolds}], these resolutions also correspond to $\Delta_p^+ = \Delta/l_p \approx 15$ and $\Delta_p^+ \approx 120$, respectively. It is emphasized based on \citet{agrawal2023reynolds} that $\Delta_p^+ \lesssim  \mathcal{O}(10)$ is an approximate upper bound on the local grid resolution in the vicinity of the separation point to reasonably capture the onset of the flow separation. The \emph{a posteriori} sections of this article demonstrate the applicability of this resolution bound in predicting the flow separations for all considered Mach numbers in the SD-2 flow.  \\ 

\begin{table}
\centering
\begin{tabular}{ p{1cm}p{2.4cm}p{2.4cm}p{3.5cm}p{2.5cm}}
Grid & Core resolution ($\Delta/D$) & Wall resolution ($\Delta_w/D$) & Max. Pressure-gradient based wall resolution ($\Delta_w/min(l_p)$) & Total cell count (in millions) \\
\hline
$L3$  & 1/20 & 1/160 & 250 & 30    \\
$L4$ & 1/40 & 1/320  & 130 & 80   \\
$L5$  & 1/80 & 1/640  & 66 & 320   \\
$L6$ & 1/160 & 1/1280  & 30 & 750   \\
$L7$  & 1/320 & 1/2560  & 15 & 3000 \\
\end{tabular}
\caption{The variation in the number of points per the diameter of the exit plane, the resolution of the core region and the wall region, the total number of control volumes (CVs) for the flow inside the SD-2 diffuser. Note that $D$ is the diameter of the AIP.}
\label{table:sd2grids}
\end{table} 

\begin{figure}[!ht]
\centering
(a) \includegraphics[width=0.42\textwidth]{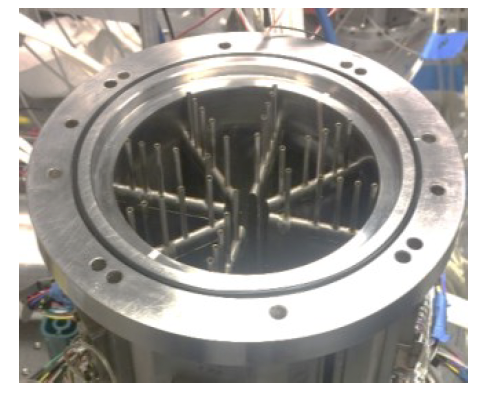}
(b)\includegraphics[width=0.41\textwidth]{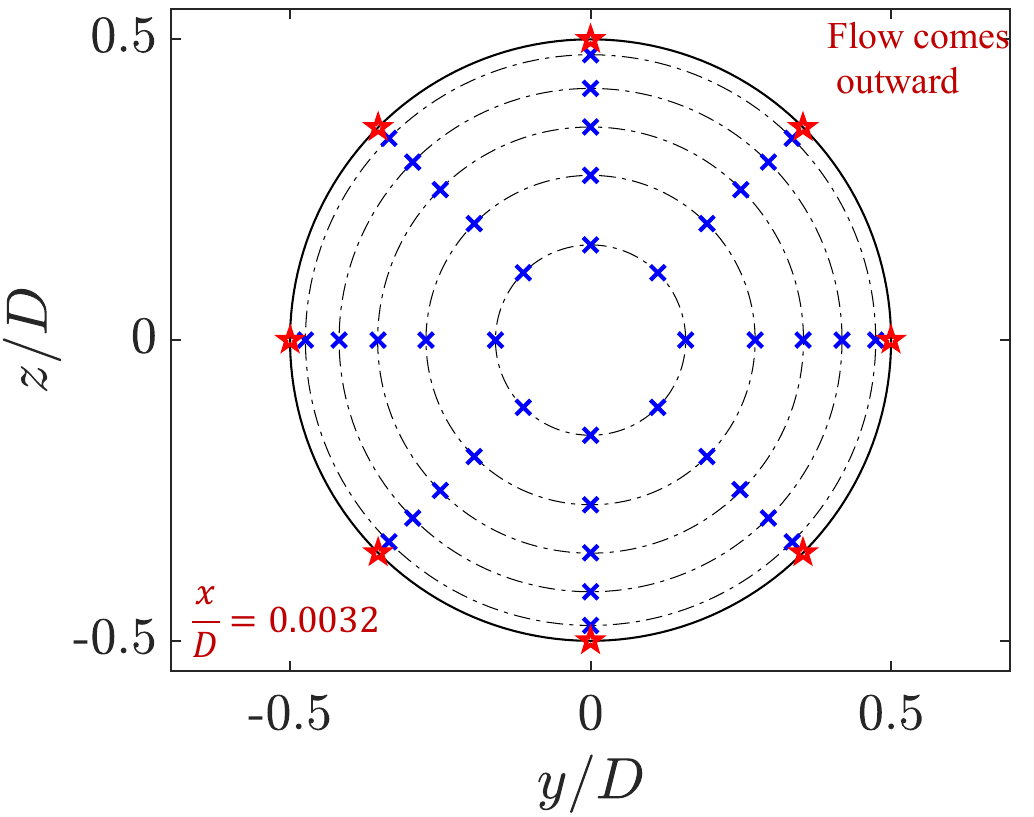}
    \caption{ (a) A top-down view of the SAE standard, pitot tube--based total pressure rakes placed at the AIP in the experiments [image reproduced from \citet{burrows2020evolution}]. (b) The positions of the 40 dynamic pressure probes (blue crosses) and the eight static pressure probes placed at the AIP to compute the Mach number and the mass flow rate inside the diffuser. In this view, $y$ and $z$ correspond to the spanwise and the vertical directions, respectively.    }
    \label{fig:40pointprobe}
\end{figure}

\noindent
The boundary conditions were set as follows: the reference total inlet pressure and the total temperature match the experimental values of $98870.8 \, \mathrm{Pa}$ and $292.2 \, \mathrm{K}$, respectively. All tunnel walls were treated viscously with a wall model.  There is 
no special treatment for transition within the tunnel bellmouth, 
and although not shown, disable the wall model in the bellmouth produced no discernible differences in the results. A characteristic mass outflow boundary condition is employed (the outlet pressure varies to hold the mass flow constant over the convective time scale of the diffuser). \citet{burrows2020evolution} experimentally measured approximate mass flow rate through a collection of a 40-point pitot rake [see Figure \ref{fig:40pointprobe}(a); the reader may refer to report \#ARP1420-B of the SAE standards for more details], placed slightly downstream of the AIP at $x/D \approx 0.003$ (the origin is at the AIP). These 40 probes [blue crosses in Figure \ref{fig:40pointprobe}(b)] and the 8-point static pressures (red pentagrams) on the sidewalls of the AIP are used to determine a Mach number, which is then used to calculate an approximate flow rate (denoted as $W_{40}$). In this work, three mass flow rates, $W_{40} = \{1.61, \, 1.95, \, \mathrm{and} \, 2.27 \} \, \mathrm{kg/s}$ are considered, which result in AIP Mach numbers, $Ma_{AIP} \approx \{0.36, \, 0.46, \mathrm{and} \, 0.54 \} $ respectively (in the experiments). \\

\section{ \texorpdfstring{WMLES results for AIP condition: $Ma_{AIP} \sim 0.54 $}{}}
\label{sec:ma0p70wmles}
\subsection{Dynamic Smagorinsky subgrid-scale and equilibrium wall models}

\begin{figure}[!ht]
\centering
    (a){\includegraphics[width=0.45\textwidth]{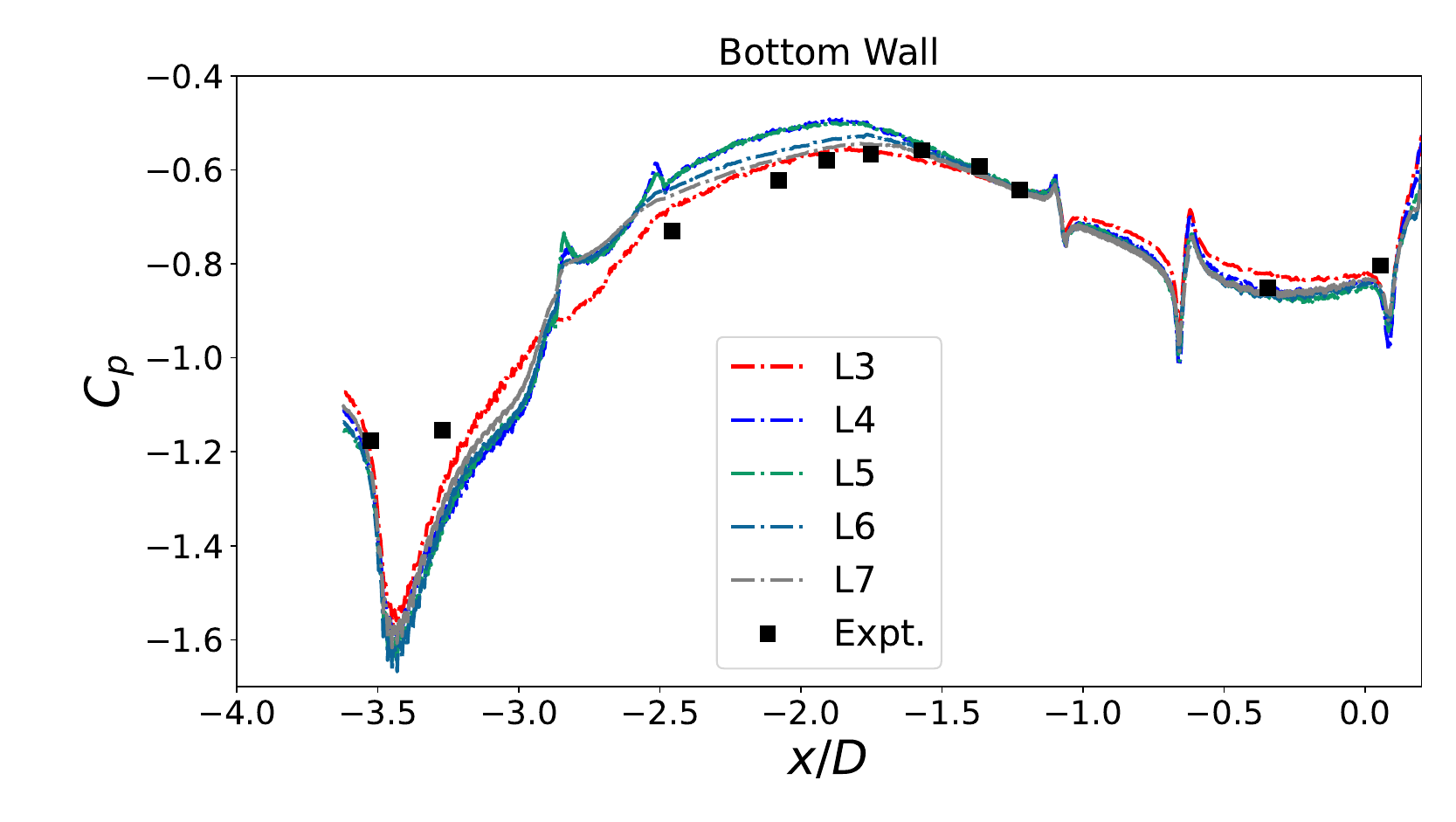}}
    (b) {\includegraphics[width=0.45\textwidth]{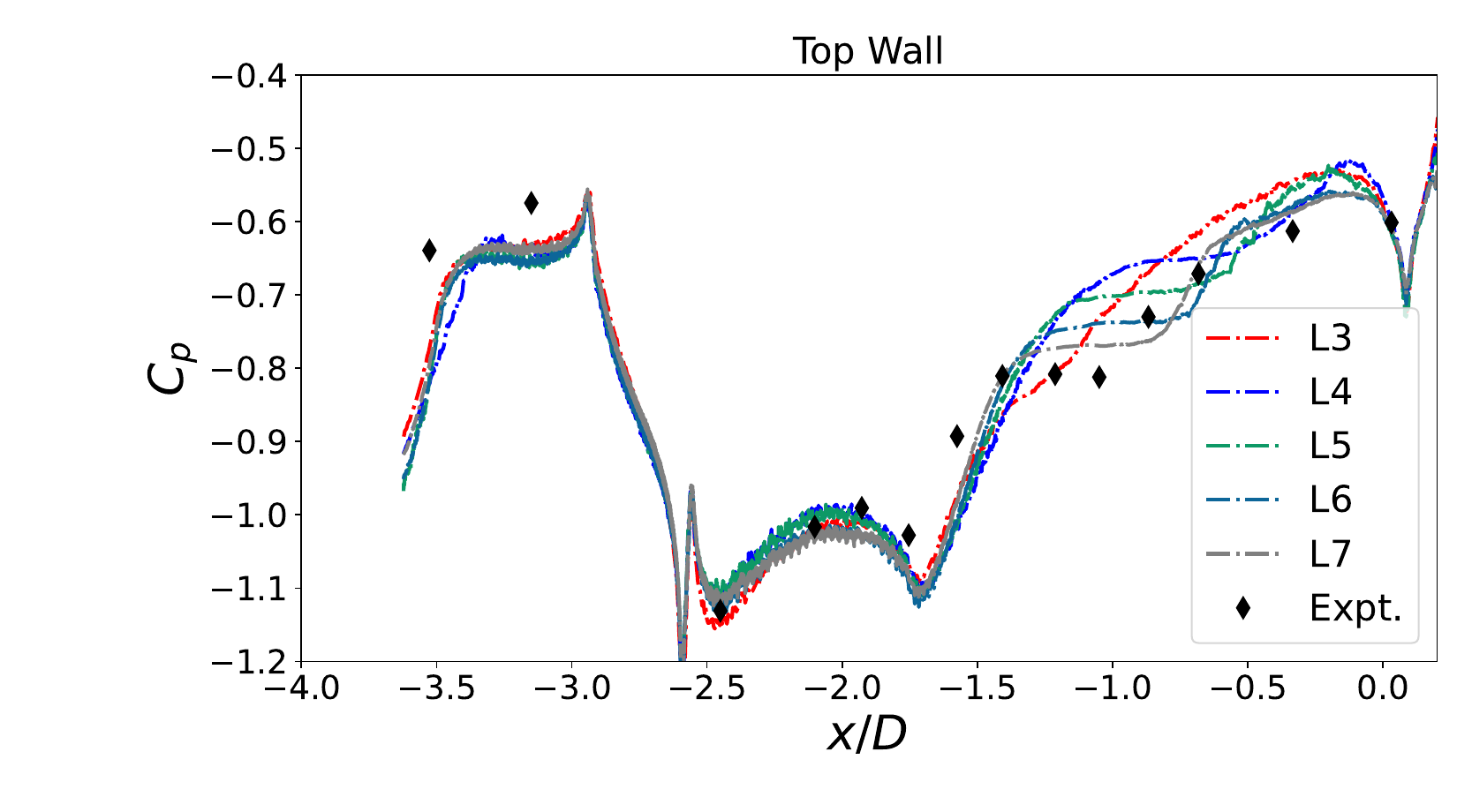}}
        
    \caption{ The prediction of surface pressure, $C_p$ (using dynamic Smagorinsky subgrid-scale and equilibrium wall models), on (a) the bottom and (b) the top walls (at the midspan) of the SD-2 diffuser of \citet{burrows2020evolution}. The flow conditions simulated achieve a transonic Mach number, $Ma = 0.7$, at the diffuser throat ($Ma_{AIP} \sim 0.54$).  }
    \label{fig:sd2dsmcp}
\end{figure}

Figure \ref{fig:sd2dsmcp} shows the distribution of the mean surface pressure coefficient ($C_p = 2(p-p^0_t)/(\gamma Ma_{t} p^0_t), $ where $p^0_t, \, Ma_{t}, \, \gamma(=1.4)$ are the reference inlet total pressure, achieved Mach number at the throat, and the specific heat ratio respectively) at the midspan on both walls. Similar to several previous demonstrations \citep{agrawal2022non,goctransoniccrm,hlcrmkonrad} in boundary layers exhibiting smooth-body separation, the DSM/EQWM models exhibit a non-monotonic convergence behavior toward the reference data upon grid refinement. On the coarsest grid, $L3$, the predicted $C_p$ agrees more with the experiments than those on $L4-L6$ grids. On the finest grid, $L7$, the $C_p$ prediction on the top wall are in better agreement with the experiments relative to the $L3$ grid. The separation 
region on both walls appear to be slightly underpredicted even on the $L7$ grid 
relative to the experiment.\\

\begin{figure}[!ht]
\includegraphics[width=1\textwidth]{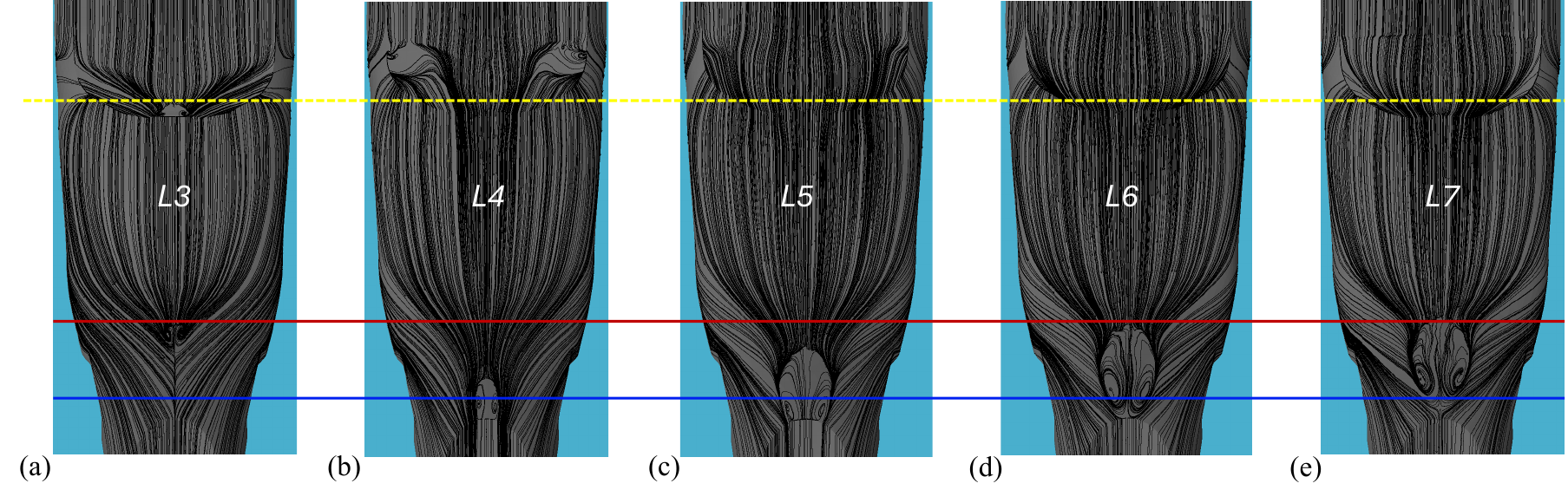}  
    \caption{ Mean near-wall streamlines on the top wall across the five grid resolutions while invoking the DSM/EQWM closures. The red and blue lines mark the mean separation and reattachment points on the $L7$ grid, respectively. The yellow dotted line marks the first turn on the top wall. The mean attached flow is directed from the top to the bottom.  }
    \label{fig:sd2dsmmeantop}
\end{figure}

\begin{figure}[!ht]
\centering
\includegraphics[width=1\textwidth]{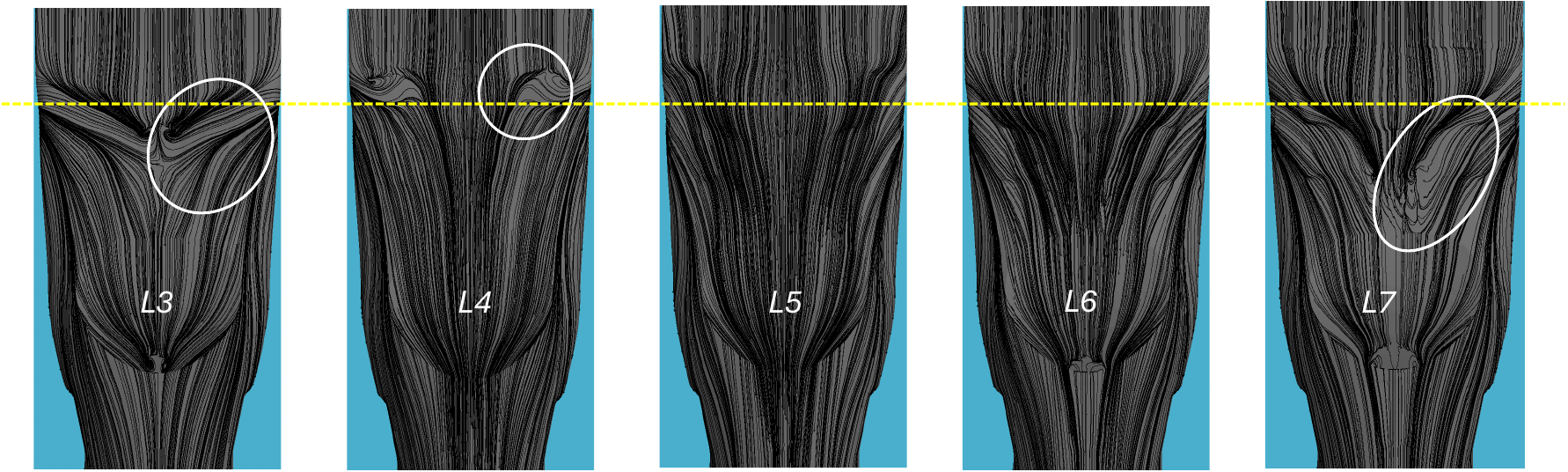}  
    \caption{Mean near-wall streamlines on the bottom wall across the five grid resolutions while invoking the DSM/EQWM closures. The red and blue lines mark the mean separation and reattachment points on the $L7$ grid, respectively. The yellow dotted line marks the first turn on the top wall. The mean attached flow is directed from the top to the bottom.  }
    \label{fig:sd2dsmmeanbtm}
\end{figure}

\noindent
 Figure \ref{fig:sd2dsmmeantop} presents near-wall streamlines projected on the top wall. The flow pattern changes significantly around the first turn upon grid refinement. The acceleration of streamlines toward the midspan is expected due to the stronger velocity in the corners (resembling a wall jet phenomenon). The mean flow directionality on the $L3$, $L6$, and $L7$ grids is in agreement with this expectation. However, the mean flow on the $L4$ grid forms a localized separation pocket upstream of the first turn. The red and blue lines denote the locations of mean flow separation and reattachment on the $L7$ grid. Upon comparing this line with the coarser grids, it is evident that the separation onset moves upstream with grid refinement. Similarly, Figure \ref{fig:sd2dsmmeanbtm} shows that on the bottom wall, localized separation pockets are formed on the $L4$ grid, whereas the $L5-L6$ grids do not exhibit any separation. The $L7$ grid exhibits a 
 flow reversal downstream of the first corner with an eventual reattachment downstream.   \\

\subsection{Dynamic tensor coefficient subgrid-scale and sensor wall models}

 \begin{figure}[!ht] 
\centering
   (a) {\includegraphics[width=0.43\textwidth]{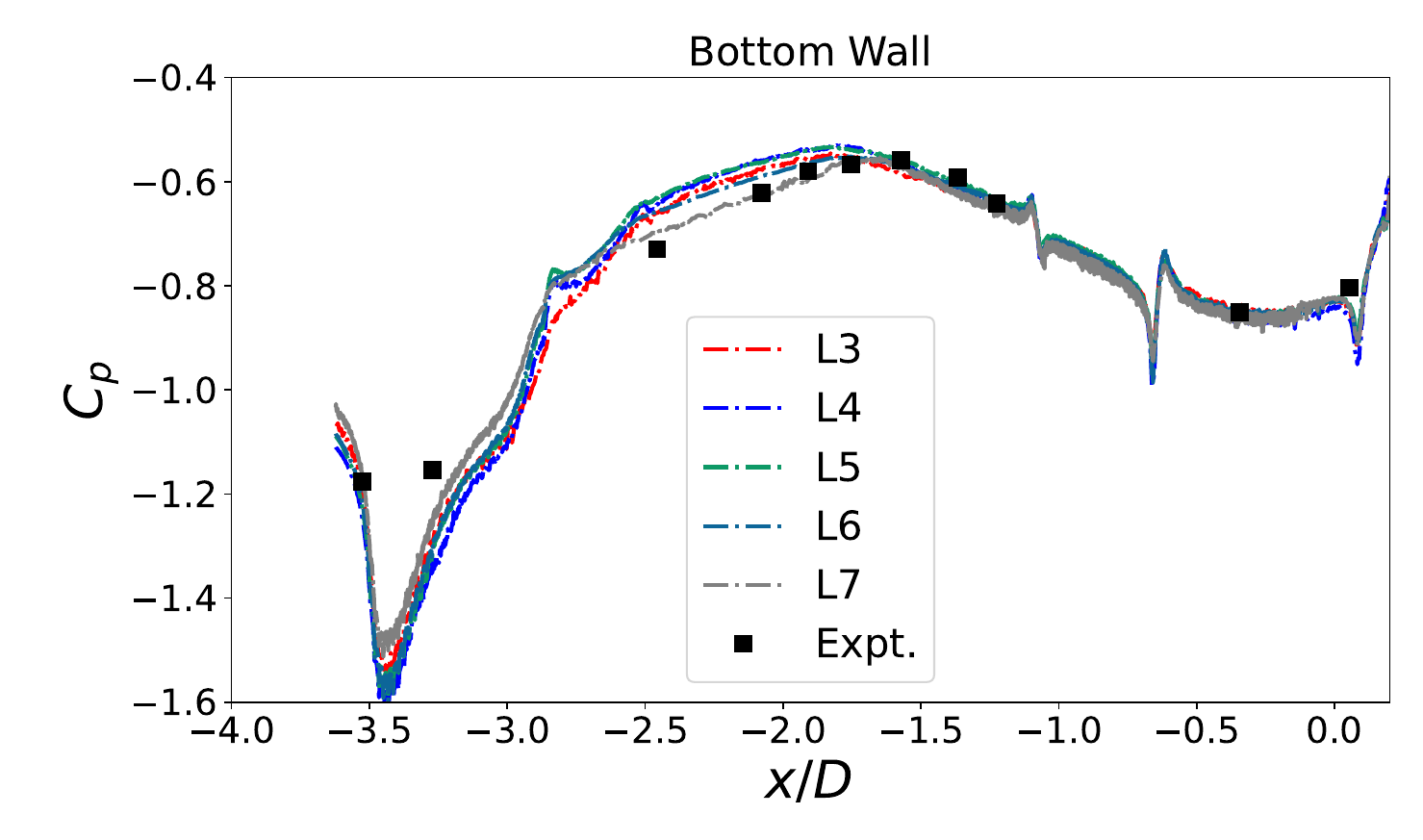}} 
   (b){\includegraphics[width=0.43\textwidth]{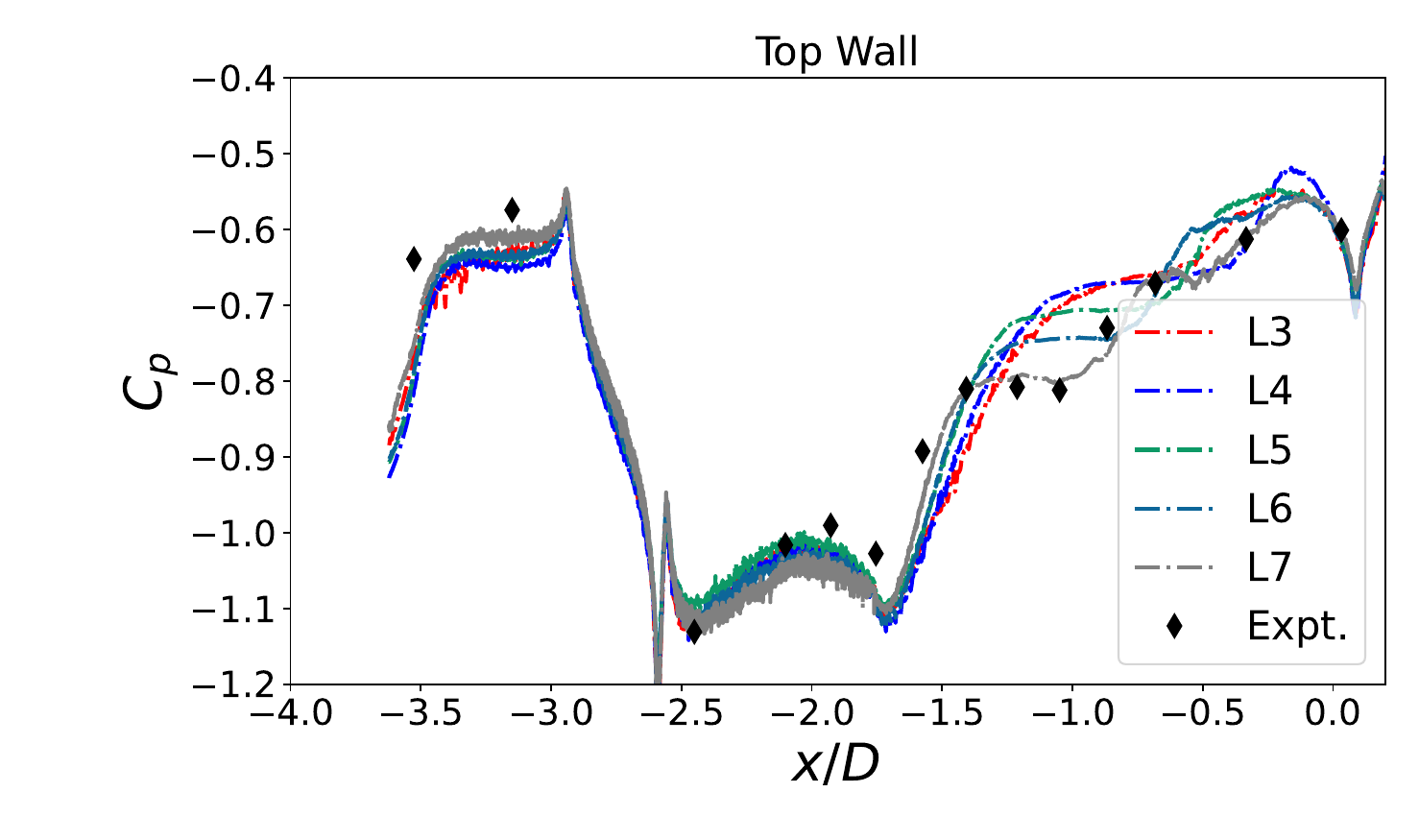}} 
   \caption{ The prediction of surface pressure, $C_p$ (using DTCSM/SWM models), on (a) the bottom and (b) the top walls (at the midspan) of the SD-2 diffuser of \citet{burrows2020evolution}. The flow conditions simulated achieve a transonic Mach number, $Ma = 0.7$, at the diffuser throat ($Ma_{AIP} \sim 0.54$).  }
     \label{fig:sd2dtcsmswmcp}
 \end{figure}

\begin{figure} 
\centering
\includegraphics[width=1\textwidth]{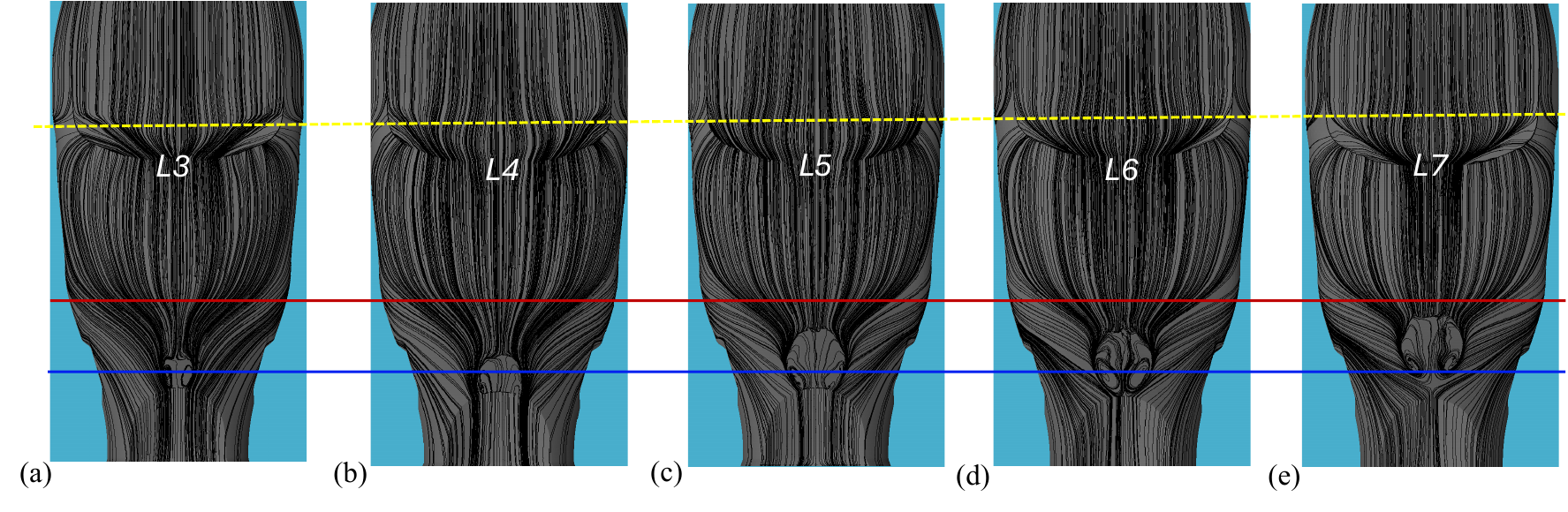}
\caption{ Mean near-wall streamlines on the top wall across the five grid resolutions while invoking the DTCSM/SWM closures. The red and blue lines mark the mean separation and reattachment points on the $L7$ grid, respectively. The yellow dotted line marks the first turn on the top wall. The mean attached flow is directed from the top to the bottom.   }
    \label{fig:sd2dtcsmswmmeantop}
\end{figure}

\begin{figure} 
\centering
\includegraphics[width=1\textwidth]{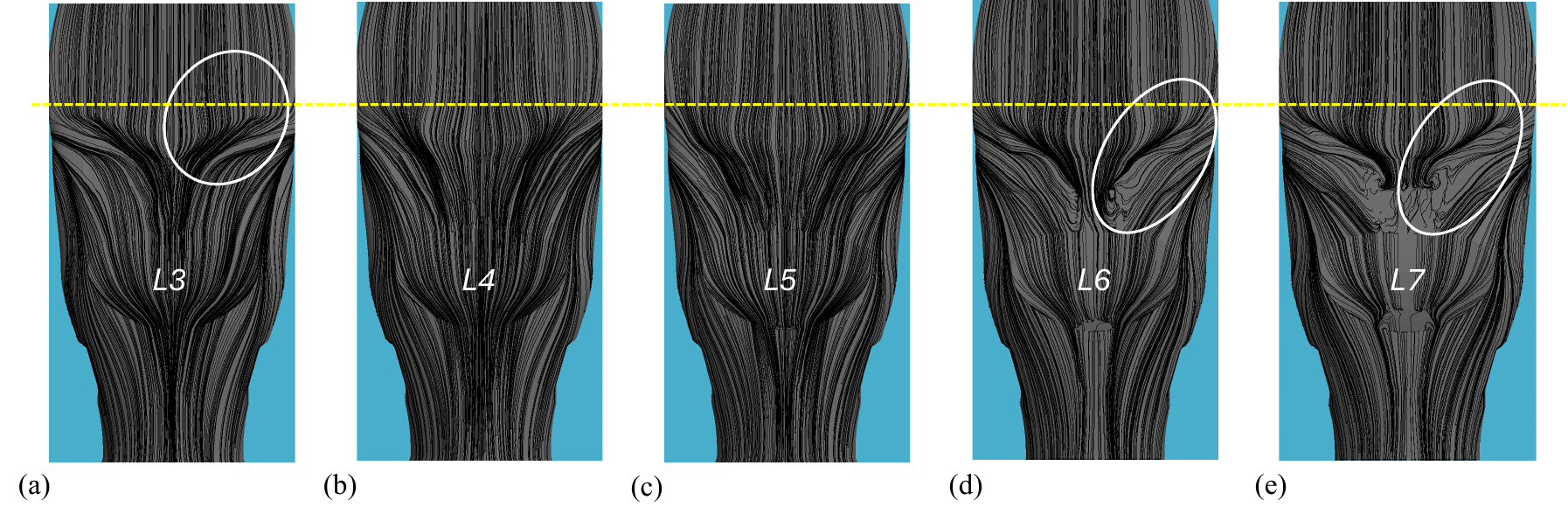}    
\caption{ Mean near-wall streamlines on the bottom wall across the five grid resolutions while invoking the DTCSM/SWM closures. The red and blue lines mark the mean separation and reattachment points on the $L7$ grid, respectively. The yellow dotted line marks the first turn on the top wall. The mean attached flow is directed from the top to the bottom.  }
    \label{fig:sd2dtcsmswmmeanbtm}
\end{figure}

Figure \ref{fig:sd2dtcsmswmcp} compares the grid convergence of $C_p$ at the midspan using the DTCSM/SWM models. In line with \citet{agrawal2022non,agrawal2024non}, the predictions from DTCSM/SWM generally monotonically approach the experiments upon grid refinement. On the finest grid, $L7$, where $\Delta/min(l_p) \sim \mathcal{O}(10)$, the predictions agree well with the experiments.  Theoretical error analysis of wall
models in smooth body separating flow
suggested that $\Delta/l_p \lesssim \mathcal{O}(10)$ was necessary for sufficiently small errors in the surface pressure coefficient \cite{agrawal2023grid} . Figures \ref{fig:sd2dtcsmswmmeantop} and \ref{fig:sd2dtcsmswmmeanbtm} show the near-wall streamlines on the top and bottom walls, respectively. The flow separation on the bottom wall appears only on the two finest grids ($L6$ and $L7$), marked by the inward-moving corner flow (within the white ovals). In comparison to the DSM/EQWM predictions, the acceleration of the corner flow toward the centerline is consistently observed on all grids, and no local separation pockets are observed on the first turn. The separation point on the second turn and its consequent reattachment move upstream with grid refinement with the two finest grids producing nearly grid converged results.  \\

\subsection{Activity of the non-equilibrium wall model sensor}
\begin{figure}[!ht] 
\centering
    {\includegraphics[width=0.5\textwidth]{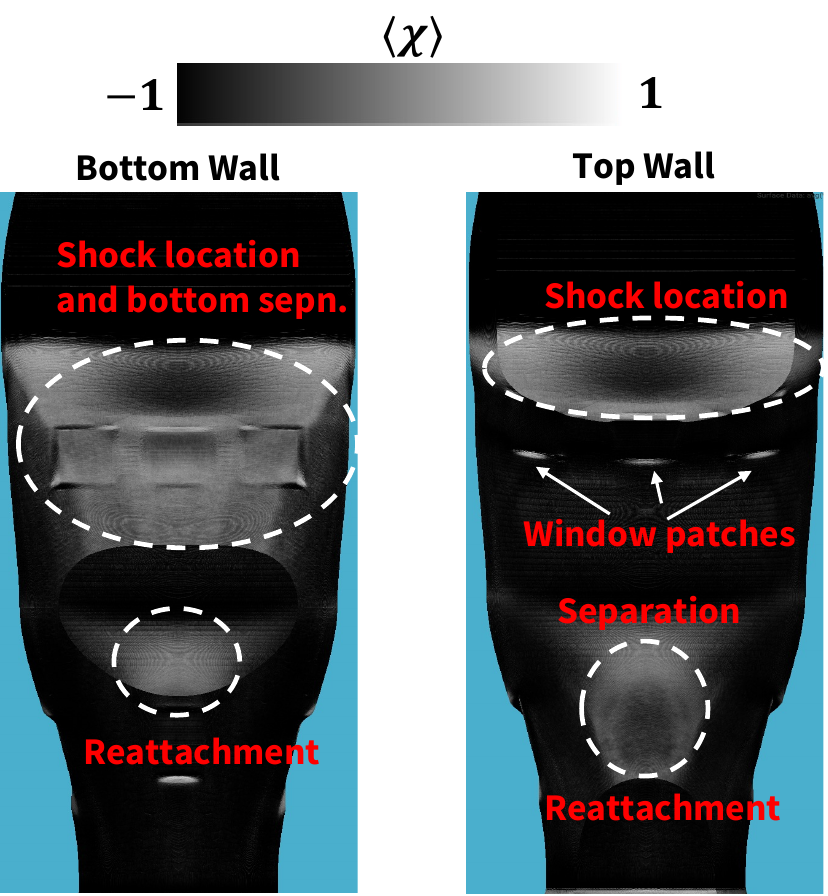}}
    \caption{ Contour of the time-averaged activity of the proposed sensor for the $L5$ grid, as measured by $\langle \chi \rangle $, where, if $\langle \chi \rangle \geq 0$, the sensor is active in the mean and vice versa. Although not shown, the sensor activity contours are qualitatively similar on other grids. }
     \label{fig:sd2L5bfsviz}
 \end{figure}

As detailed in \citet{agrawal2024non}, the sensor aided non-equilibrium wall closure attempts to capture the flow regions in the vicinity of a separation bubble induced due to mild adverse pressure gradients. Figure \ref{fig:sd2L5bfsviz} presents the time-averaged activity of this sensor (denoted as $ \langle \chi \rangle$, where $\langle \cdot \rangle$ is the time averaging operator). Note that $\langle \chi \rangle \geq 0$ corresponds to a separation bubble.  The sensor identifies non-equilibrium effects near the shock, separation, and reattachment points on both walls. Further, the sensor is first active (white region) on the corners on the bottom walls, which eventually extends toward the centerline, consistent with the experimental observations of the flow separation emanating from the corners \citep{burrows2020evolution}.   \\

\subsection{Error Convergence with Grid Refinement }
\begin{figure}[!ht]
\centering
    {\includegraphics[width=0.5\textwidth]{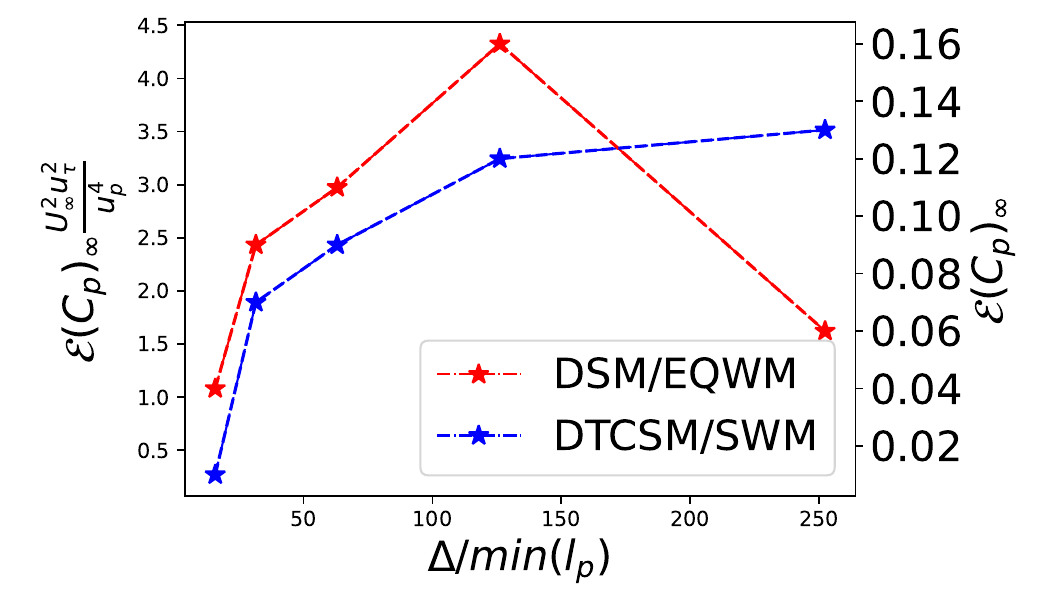}}
    \caption{ The dependence of the maximal error ($L_\infty$ norm) of the group, of the non-dimensional group,  $E(C_p)_{\infty} \frac{u^2_\tau U^2_\infty}{u^4_p}$,  and of $E(C_p)_{\infty}$, on grid resolution, $\Delta/min(l_p)$ for the flow inside the SD-2 diffuser. The two error profiles correspond to the LES predictions from the DSM/EQWM and DTCSM/SWM closure models.   }
     \label{fig:sd2dsmdtcsmerrorcompare}
 \end{figure}

 \noindent
The error convergence rate (as measured through the maximal error in $C_p$) on the top wall is also compared between the two modeling combinations in Figure \ref{fig:sd2dsmdtcsmerrorcompare}. Both DSM/EQWM and DTCSM/SWM exhibit a slow convergence toward the reference experiments. For the latter models, the error nearly asymptotes to a constant value as the predicted $C_p$ on these 
coarse grids approaches the pressure 
distribution of an inviscid flow. On the finer grids, per \citet{agrawal2023grid}, the errors decay with $\Delta_{p}^{+}$. This is found to hold for both DSM/EQWM and DTCSM/SWM models. \\

\subsection{Predictions at the Aerodynamic Interface Plane}
\noindent
The predictions of the mean flow recovery and azimuthal flow distortion at the AIP are compared with experimental data are 
compared below. A discussion of maximum azimuthal flow distortion is provided in Section \ref{sec:machsens}. Note that in the subsequent discussion, the following data-filtering operations are performed to the simulation data (for consistency with the experiments). The initial, transient data (first 20\% of the time signal) is removed from consideration. Next, the time-series from of the total pressure from the 40 dynamic probes is low-pass filtered using a sixth-order Butterworth filter with a cutoff frequency equal to 1500 Hz. \\

\noindent
On the finest grid, $L7$ (where the mean $C_p$ predictions are reasonably agreeable with the experiments), the simulations are run for $\approx 30 \times 10^{-3} \mathrm{sec}$, roughly equivalent to $ \sim 15$ flow passes (one flow pass is equal to $L_x/U_\infty$ where $L_x$ is the streamwise length of the domain and $U_\infty$ is the centerline velocity of the flow at the throat). The experiments were recorded for $\approx 5 \, \mathrm{sec}$, which is two orders of magnitude larger than the integration time on the $L7$ grid.

\subsubsection{Flow recovery }
\begin{figure}[!ht]
\centering
  {\includegraphics[width=0.9\textwidth]{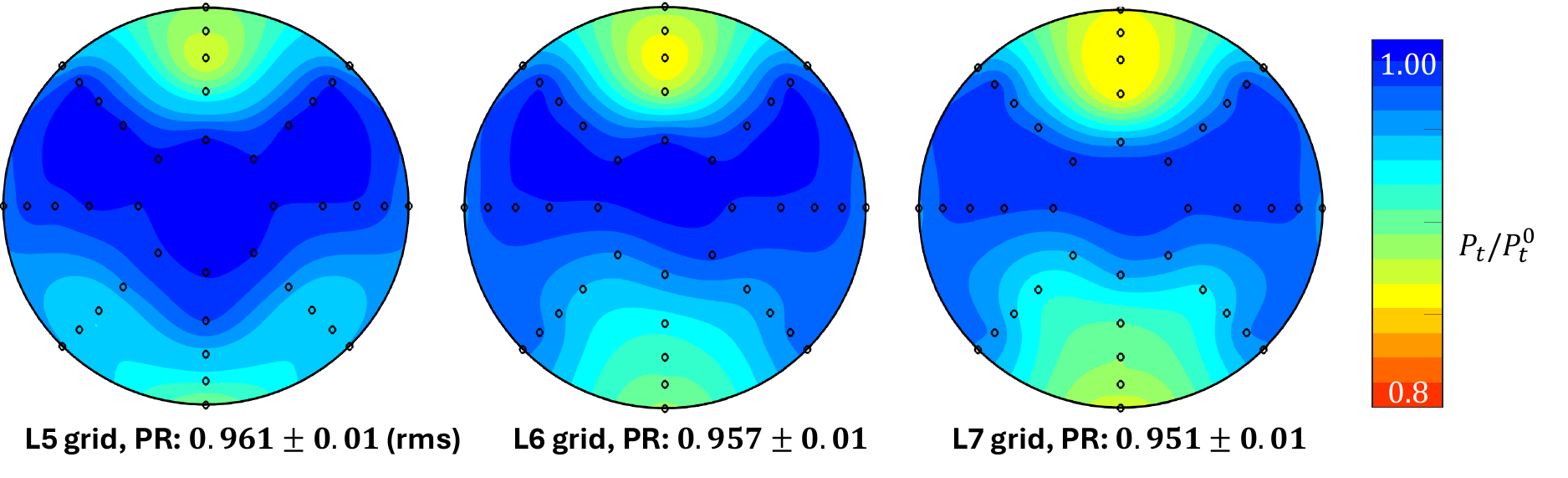}}
    \caption{Contours of the averaged total pressure measured at the 40 probes placed at the AIP on the $L5$, $L6$ and $L7$ grids with DTCSM/SWM models. The black dots in the simulation contours correspond to the 40 probe locations. $P^0_t$ is the reference total pressure at the inlet of the diffuser domain.}
     \label{fig:sd2aipdtcsmgridpressure}
 \end{figure}

The pressure recovery ($\mathrm{PR}$) at the AIP (for the $Ma_{AIP} \sim 0.54$ flow) is the ratio of the face-averaged total pressure at the exit to the inlet pressure. Figure \ref{fig:sd2aipdtcsmgridpressure} shows the predictions of $\mathrm{PR}$ for 
various grid resolutions. Only the finest three grids from the DTCSM/SWM models are considered for brevity. As the grid is refined, a consistent top wall pressure loss due to the secondary vortices generated by the separation is observed. The loss at the top center becomes stronger with grid refinement as the extent of the flow separation increases. In contrast, the pressure loss changes significantly on the bottom wall as the bottom wall separation is only observed on the $L6$ and $L7$ grids. The larger separation on the finer grids also decrease the net $\mathrm{PR}$ from $0.961$ on the $L5$ grid to $0.951$ on the $L7$ grid.  \\

\begin{figure}[!ht]
\centering
    {\includegraphics[width=0.9\textwidth]{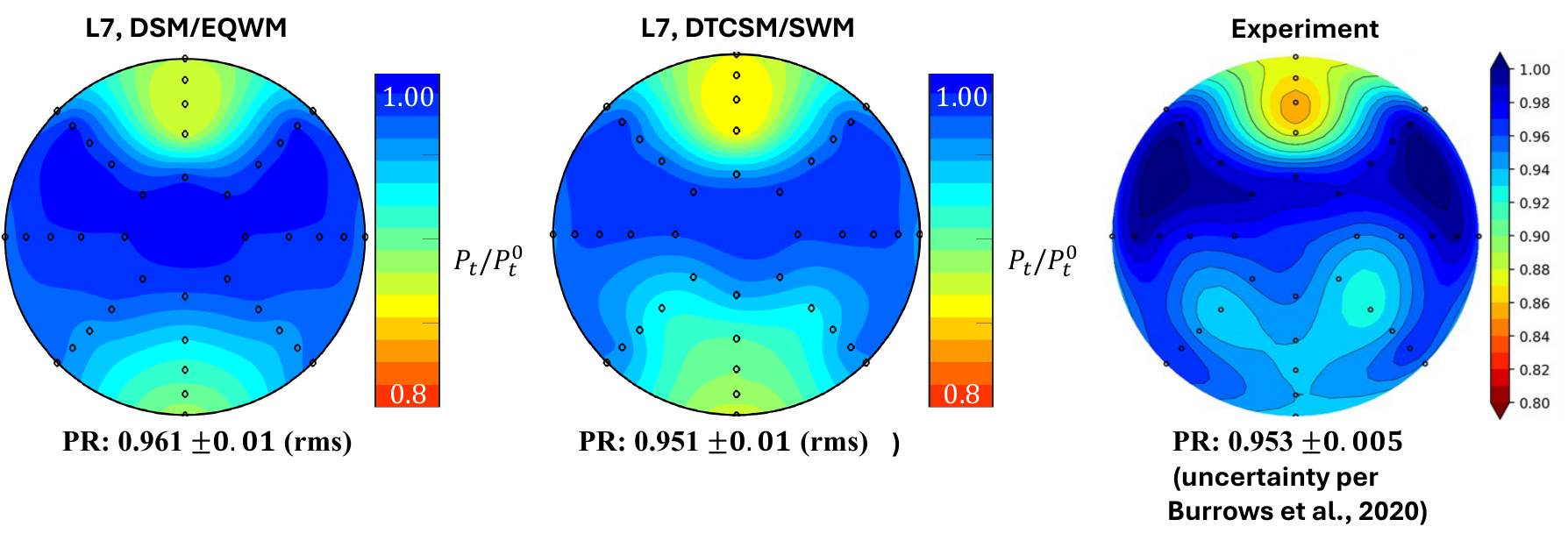}}
    \caption{  The comparison between the contours of the averaged total pressure measured at the 40 probes placed at the AIP on the $L7$ grid with the DSM/EQWM,  DTCSM/SWM models and the experiments \citep{burrows2020evolution}. The black dots in the simulation contours correspond to the 40 probe locations. }
     \label{fig:sd2aipdtcsmL7pressure}
 \end{figure}

\noindent
Figure \ref{fig:sd2aipdtcsmL7pressure} compares the pressure recovery on the $L7$ grid (for the $Ma_{AIP} \sim 0.54$ flow) from DSM/EQWM and DTCSM/SWM models with the experiments of \citet{lakebrink2019toward}. In comparison to the DSM/EQWM predictions, the DTCSM/SWM results provide a better agreement for 
the total pressure losses on the wall. We note that the prediction of the pressure recovery contour and 
the wall pressure coefficient from DSM/EQWM on the $L7$ grid is similar to the DTCSM/SWM results on the $L6$ grid. Quantitatively, the value of $\mathrm{PR}$ from DTCSM/SWM is within the experimental uncertainty bounds. 

\subsubsection{Face-averaged distortion}
The face-averaged azimuthal distortion (denoted as $\mathrm{dPcP}$) describes the azimuthal pressure losses at the five dynamic pressure rings (see Figure \ref{fig:40pointprobe}). A simplified mathematical representation of $\mathrm{dPcP}$ is given as
\begin{equation}
    \mathrm{dPcP} \equiv \sum^{\mathrm{Number \, of \, rings}}_{i=1}  \mathrm{dPcP}_i = \sum^{\mathrm{Number \, of \, rings}}_{i=1}  \bigg[ \frac{ \mathrm{PAV} - \mathrm{PAV_{low}}}{\mathrm{PAV}} \bigg]_i ,
    \label{eqn:dpcpeqn}
\end{equation}
where $\mathrm{PAV}_i$ is the ring-averaged total pressure and $\mathrm{PAV_{low, \, i}}$ is the ring-averaged pressure in the low pressure region (the azimuthal extent of the ring where the pressure falls below $\mathrm{PAV}_i$). A complete description of $\mathrm{dPcP}$ can be found in report \#ARP1420-B of the SAE standards.\\

\begin{figure}[!ht]
\centering
    {\includegraphics[width=0.70\textwidth]{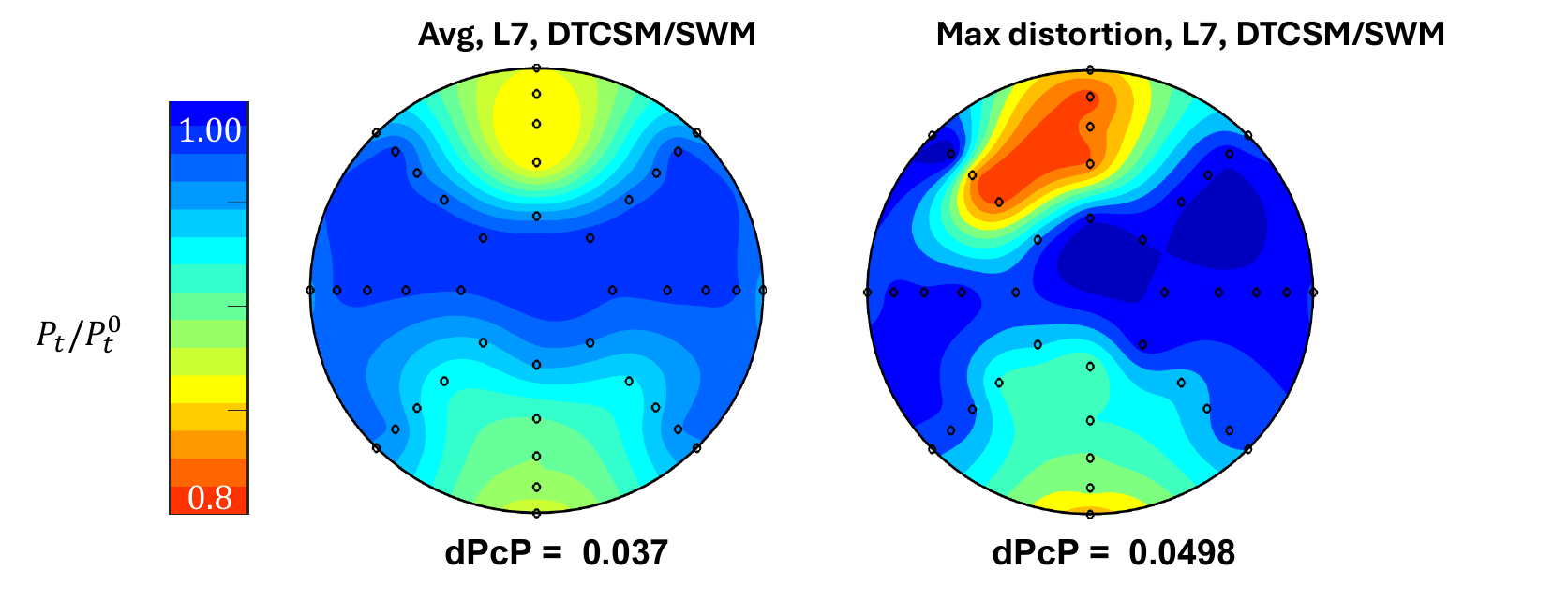}}
    \caption{ The contours of (left) mean total pressure recovery at the AIP and (right) at the instant of mean and maximum flow distortion for the flow condition: $Ma_{AIP} \sim 0.54$. }
     \label{fig:pdfdpcp}
 \end{figure}

\noindent
The predicted time-averaged dPcP on the $L7$ grid with DTCSM/SWM and DSM/EQWM is $0.037 \pm 0.005$ and $0.032 \pm 0.005$, respectively, in comparison to experimental value, $\mathrm{avg(dPcP)} = 0.038 \pm 0.002$. The variability in the simulations corresponds to the root mean square (rms) of the predicted $\mathrm{dPcP}$. The DSM/EQWM models underpredict the averaged distortion, likely due to the diminished separation on the bottom wall.  The averaged $\mathrm{dPcP}$ is within one standard deviation of the experiments for the DTCSM/SWM model. As a measure of unsteadiness in the wall-modeled LES data, Figure \ref{fig:pdfdpcp} presents the contours of the total pressure at the instant of $max(\mathrm{dPcP})$. The corresponding $\mathrm{dPcP}$ value is significantly higher than the mean value, with a more intense top wall separation, and an asymmetric (in the span) bottom wall separation.   

\section{Effect of Mach number on flow distortion}
\label{sec:machsens}

\noindent
An increase in the Mach number for a serpentine inlet generally leads to an increased flow distortion and diminished flow recovery. For instance, in transonic conditions, \citet{burrows2020evolution} observed instantaneous, large-scale structures at the AIP where the pressure losses on the top and the bottom walls intersect, forming one large region of pressure loss. On the other hand, at the lower (subsonic) throat Mach numbers, the degree of interaction between the top and the bottom wall separation is relatively diminished with less correlated spatial scales across the AIP.\\

\noindent
Thus, in this section, we compare the predictions of $C_p$, $\mathrm{PR}, \, avg(\mathrm{dPcP}), \, max(\mathrm{dPcP})$ against the experiments \citep{burrows2020evolution} for two lower Mach numbers ($Ma_{AIP} = 0.36, 0.46$). For brevity, only the solutions from $L5$-$L7$ grids are presented. It is reiterated, per \citet{agrawal2023grid}, that, theoretically, only on the $L7$ grid are the solutions expected to compare reasonably with the experiments ($\Delta \lesssim \mathcal{O}(10) min(l_p))$.

\subsection{Mean flow predictions}
\noindent

\begin{figure}[!ht] 
\centering
   (a) {\includegraphics[width=0.45\textwidth]{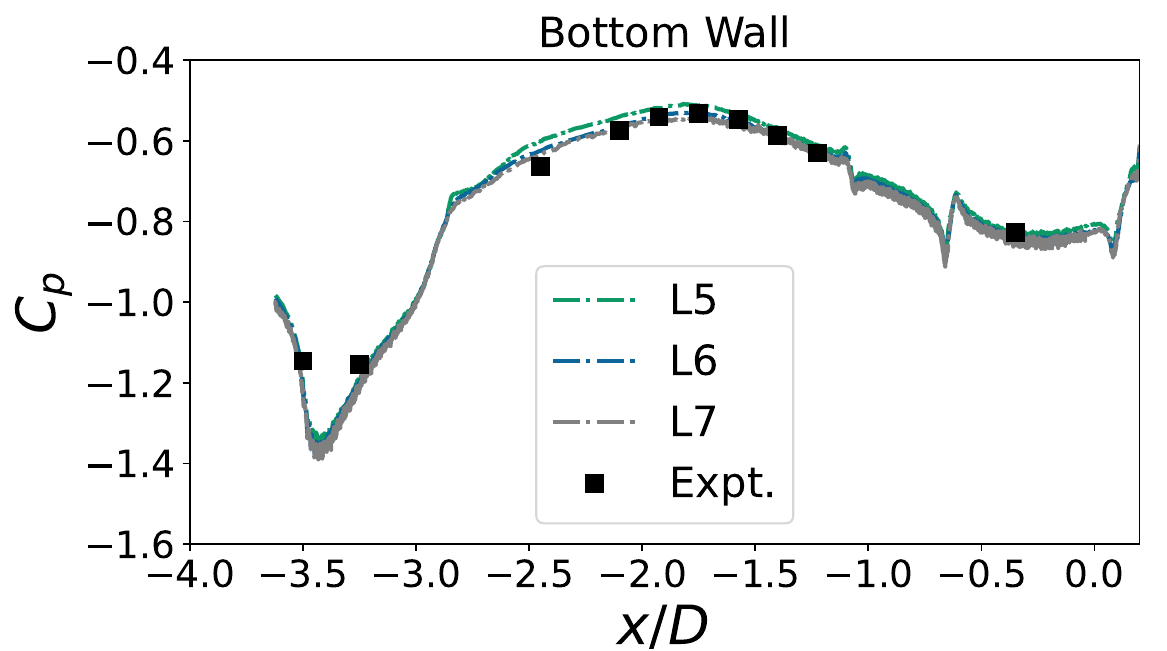}} 
   (b){\includegraphics[width=0.45\textwidth]{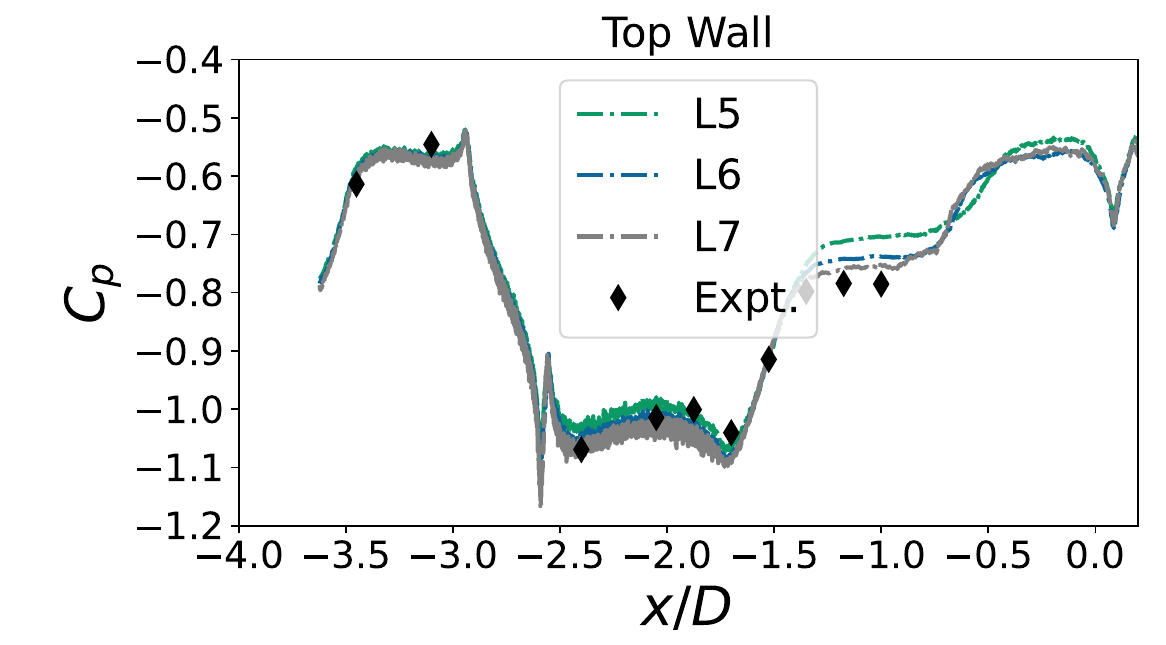}}
    \caption{ The prediction of surface pressure, $C_p$ (using DTCSM/SWM models), on (a) the bottom and (b) the top walls (at the midspan) of the SD-2 diffuser of \citet{burrows2020evolution}. The flow conditions achieve a Mach number, $Ma \sim 0.36$, at the AIP. }
     \label{fig:sd2dtcsmswmcp_3p55}
 \end{figure}

\begin{figure}[!ht] 
\centering
   (a) {\includegraphics[width=0.45\textwidth]{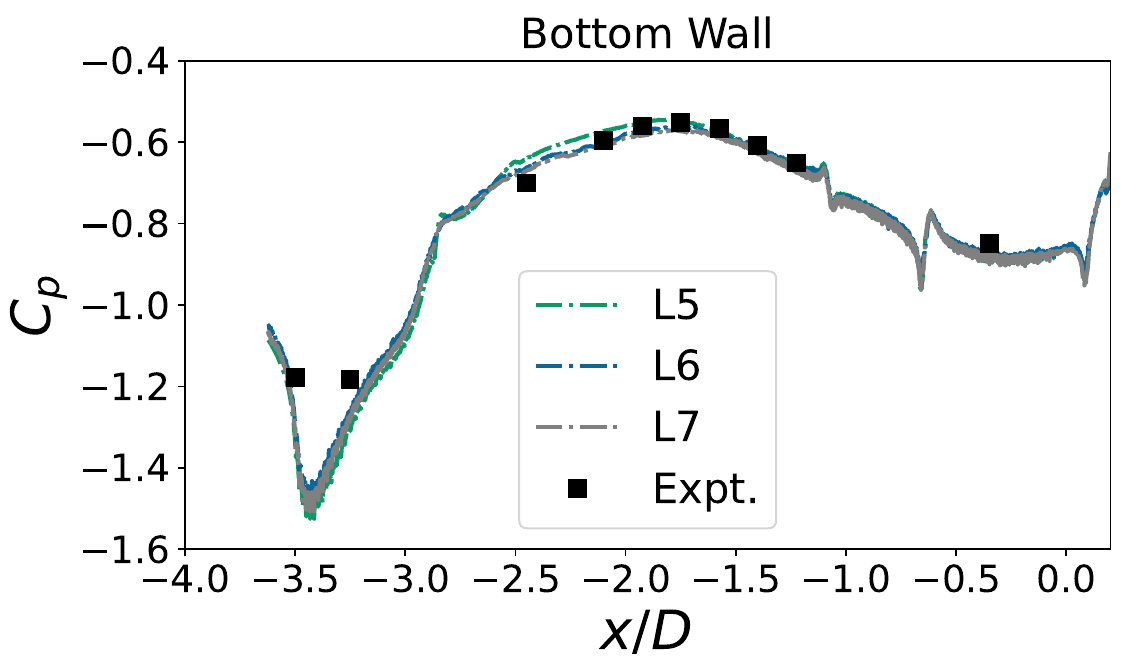}} 
   (b){\includegraphics[width=0.45\textwidth]{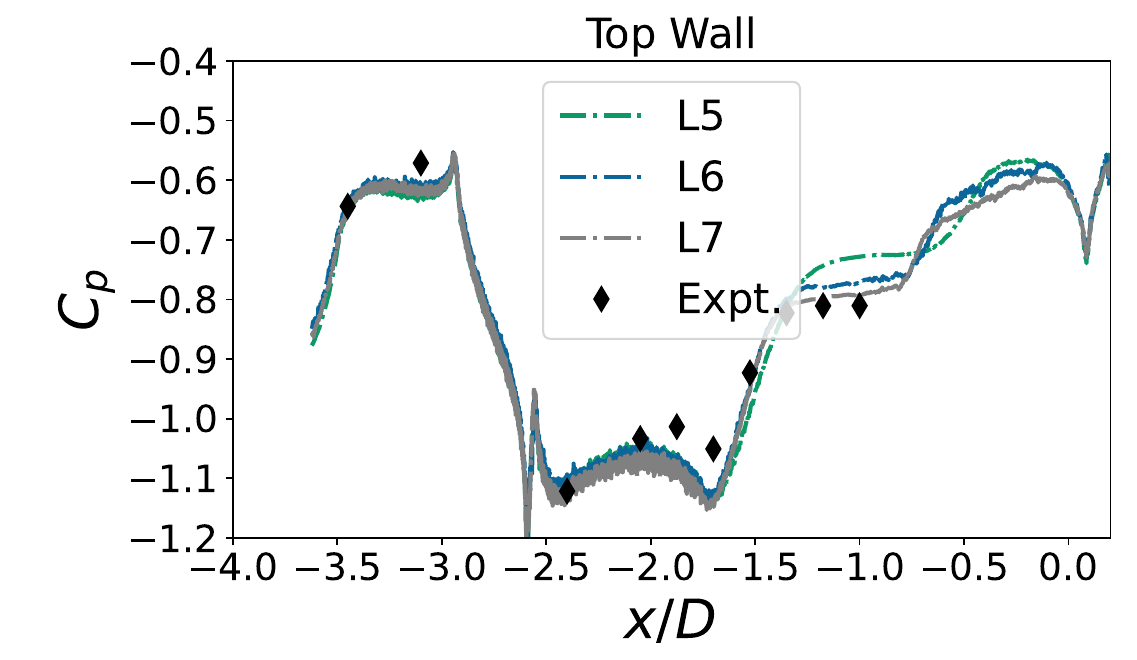}}
    \caption{ The prediction of surface pressure, $C_p$ (using DTCSM/SWM models), on (a) the bottom and (b) the top walls (at the midspan) of the SD-2 diffuser of \citet{burrows2020evolution}. The flow conditions achieve a Mach number, $Ma \sim 0.46$, at the AIP. }
     \label{fig:sd2dtcsmswmcp_4p31}
 \end{figure}

\noindent
Figures \ref{fig:sd2dtcsmswmcp_3p55} and \ref{fig:sd2dtcsmswmcp_4p31} present the predicted $C_p$ on the bottom and the top walls for the $Ma_{AIP} = \{0.36, \, 0.46 \}$ conditions, respectively. Consistent with the highest Mach number flow, the predicted solutions generally exhibit increasing flow separations upon grid refinement and on the $L7$ grid, the predicted solutions agree reasonably with the experiments. \\

\begin{figure}[!ht]
\centering
    {\includegraphics[width=0.90\textwidth]{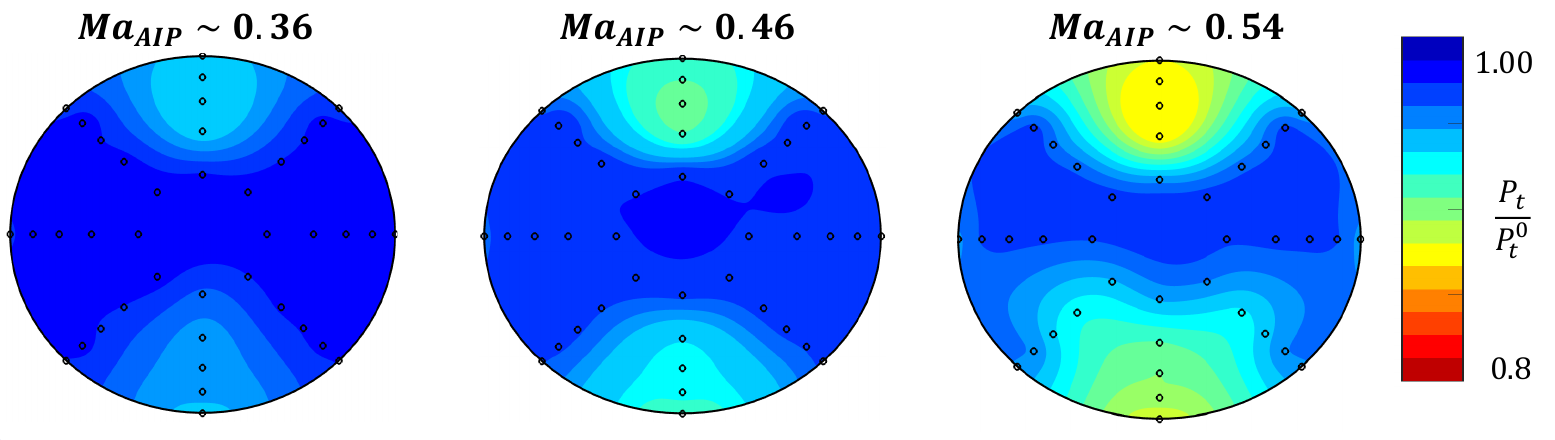}} \\
    \caption{ The time-averaged contours of total pressure recovery at the AIP corresponding to $Ma_{AIP} \sim 0.36, \, 0.46, \, 0.54$ conditions inside the SD-2 diffuser duct. }
     \label{fig:dpcp_mean_peak_mach}
 \end{figure}

\begin{figure}[!ht] 
\centering
  {\includegraphics[width=0.90\textwidth]{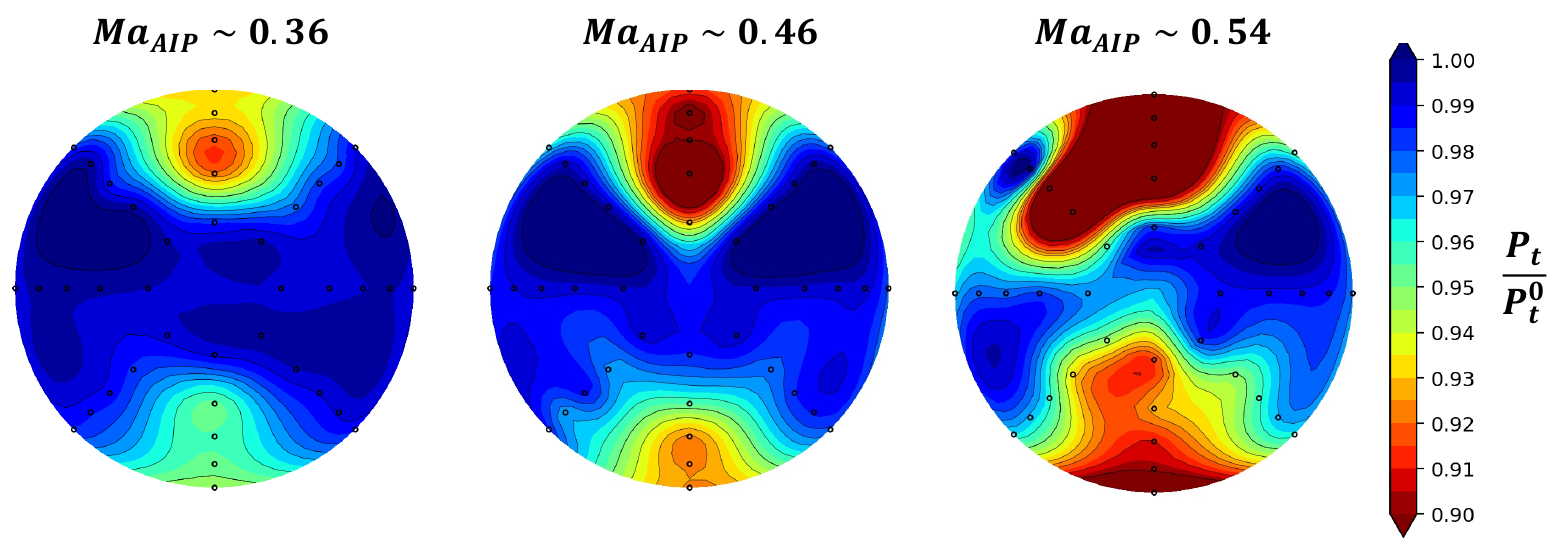}} 
    \caption{ The contours of total pressure recovery at the instance of maximum distortion corresponding to $Ma_{AIP} \sim 0.36, \, 0.46, \, 0.54$ conditions onside the SD-2 diffuser.  }
     \label{fig:maxdpcpmach}
 \end{figure}

\noindent
Figure \ref{fig:dpcp_mean_peak_mach} presents the predicted time-averaged pressure recovery contours for all computed Mach numbers. The pressure losses on the two walls increase with the Mach number. In the core, pressure loss increases at higher Mach numbers, 
which may be partly attributable to the increased dissipation of the vortices formed from the separations at higher Mach numbers. At the highest Mach number, the recovery contour exhibits a ``double lobed'' structure on the bottom wall, consistent with the presence of a counter-rotating vortex pair in the duct, formed upstream, which eventually separates from the wall. At lower Mach numbers, this flow pattern appears suppressed. A qualitative comparison of the maximum distortion ($max(\mathrm{dPcP})$) for the three flows (see Figure \ref{fig:maxdpcpmach}) suggests that the intensity of the flow interaction across the top and bottom walls becomes stronger with increasing Mach number. The peak pressure loss on the two walls also increases with Mach number, especially on the top wall, where the enhanced separation leads to a larger total pressure loss. Although not shown, the dominant spatial correlations at the AIP are of comparable to its radius, particularly at the highest Mach number, consistent with the large-scale, instantaneous interactions across the diffuser cross section.  \\ 

\begin{figure}[!ht] 
\centering
   (a) {\includegraphics[width=0.3\textwidth]{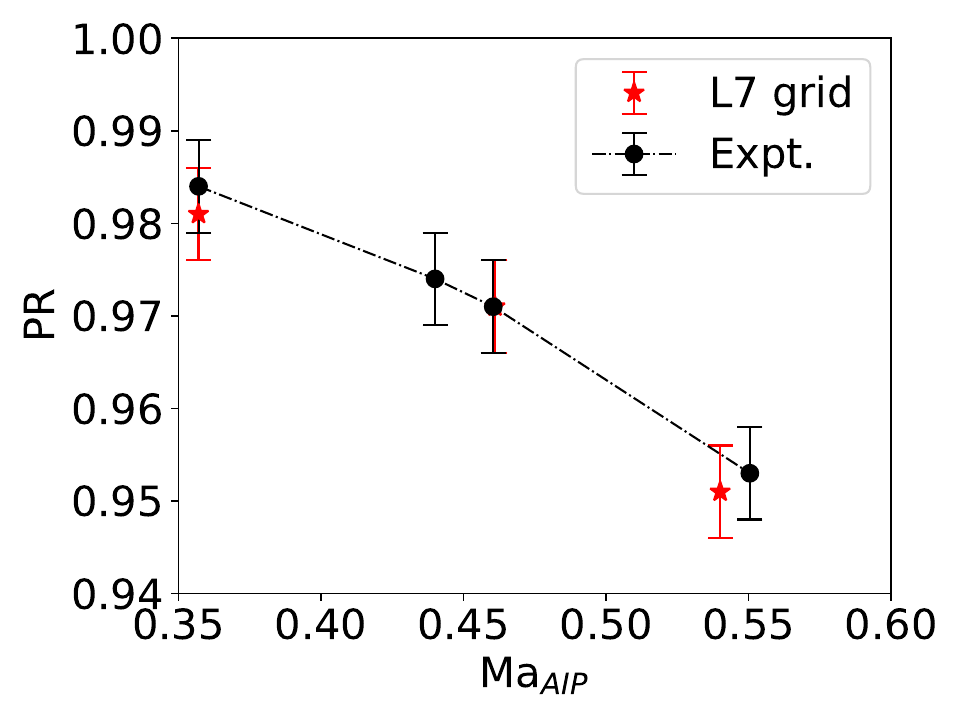}} 
   (b) {\includegraphics[width=0.3\textwidth]{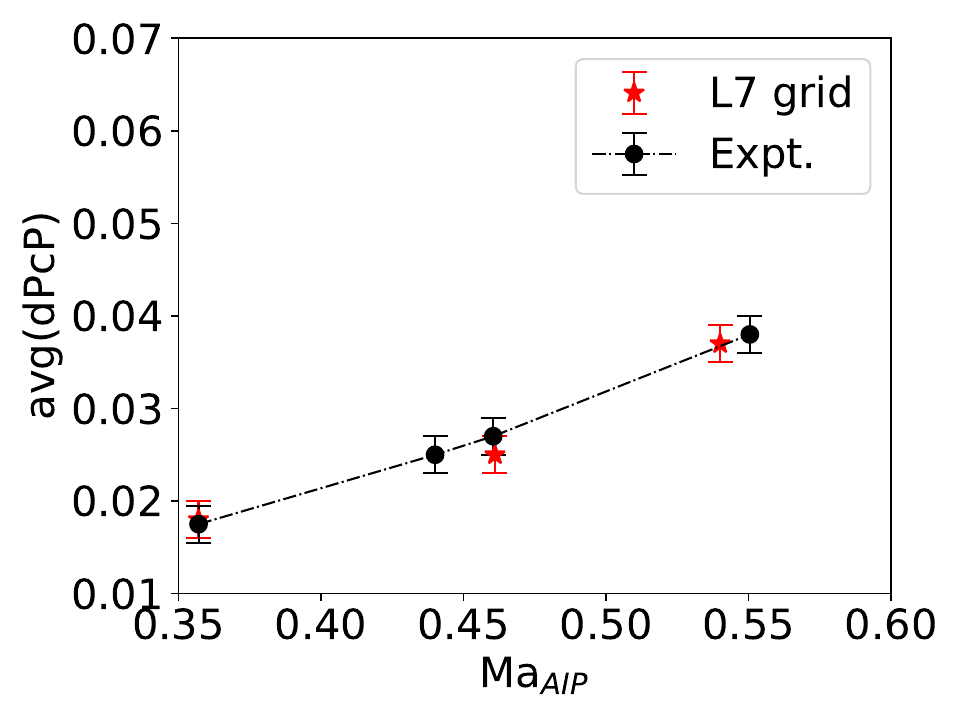}} 
   (c) {\includegraphics[width=0.29\textwidth]{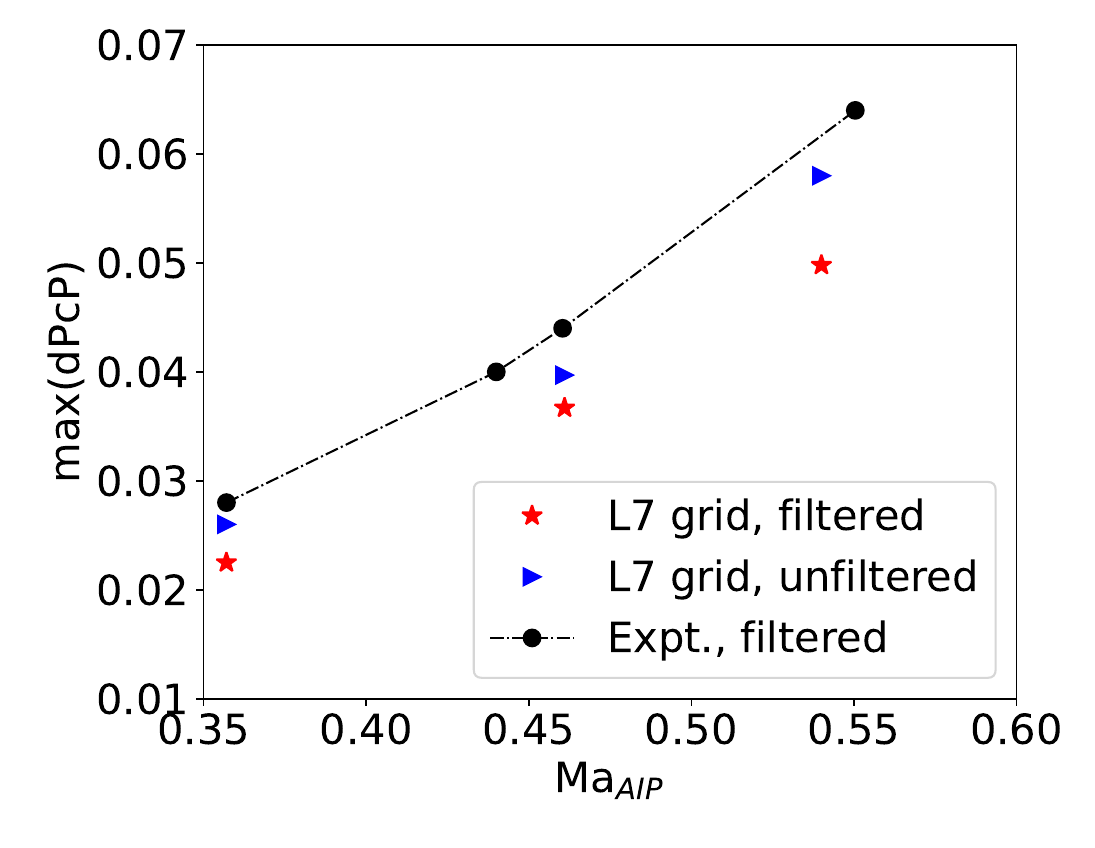}}
    \caption{ The variation in (a) mean total pressure recovery (PR) (b) time-averaged face averaged flow distortion (avg(dPcP)) and (c) maximum value of face averaged flow distortion (max(dPcP)) with Mach number for the flow inside the SD-2 diffuser.  }
     \label{fig:L7withmach}
 \end{figure}

\begin{figure}[!ht] 
\centering
  {\includegraphics[width=0.9\textwidth]{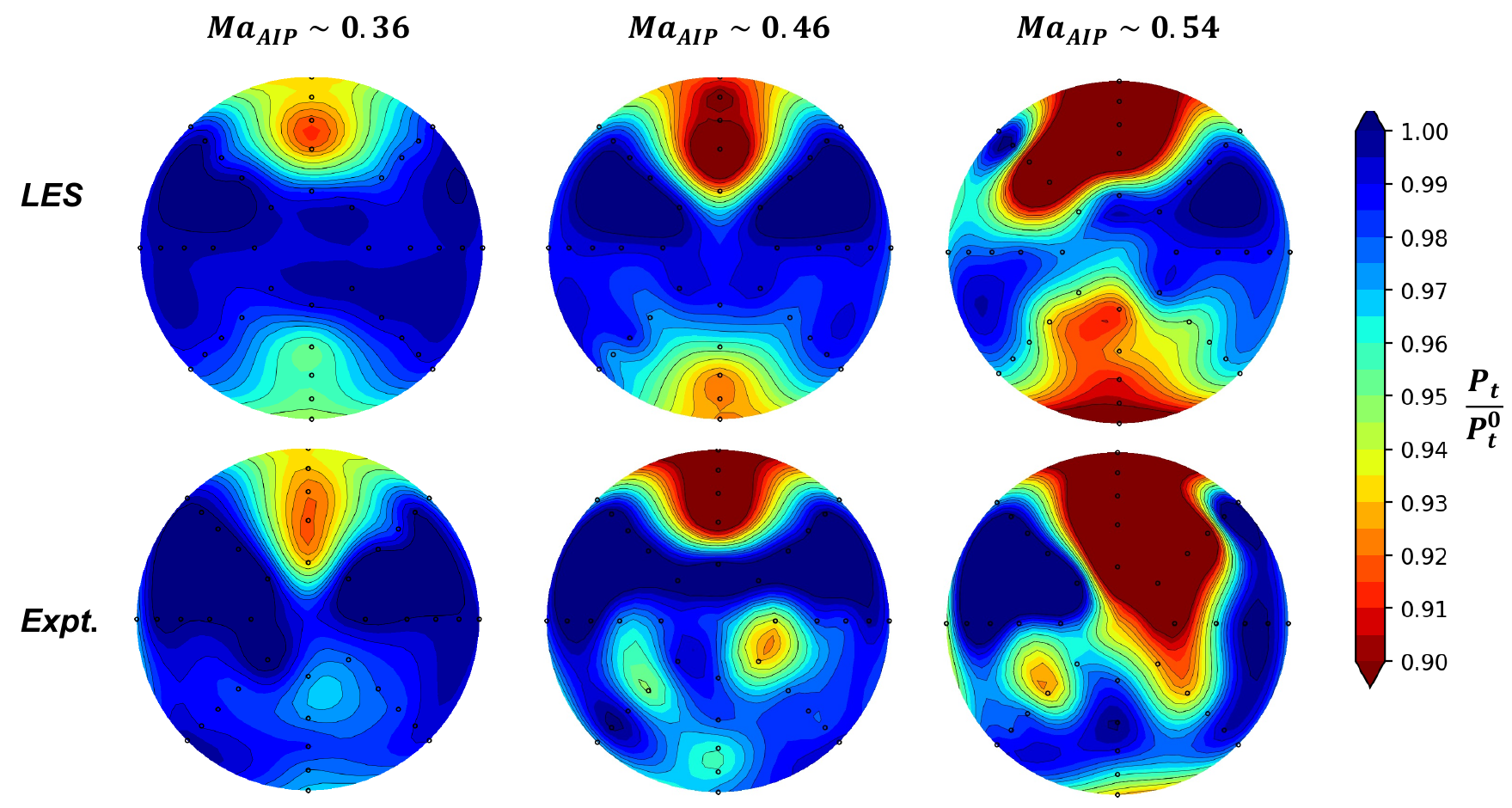}} 
    \caption{  The comparison of the contours of total pressure recovery between LES (on the $L7$ grid) and experiments at the instance of maximum distortion corresponding to $Ma_{AIP} \sim 0.36, \, 0.46, \, 0.54$ conditions onside the SD-2 diffuser.  }
     \label{fig:maxdpcpmachall}
 \end{figure}

\noindent
Figure \ref{fig:L7withmach}(a)-(b) presents the variation in the pressure recovery and azimuthal flow distortion ($\mathrm{PR}, \, avg(\mathrm{dPcP}), \, \mathrm{and} \,   max(\mathrm{dPcP})$) at the AIP as a function of Mach number. The time-averaged pressure recovery predictions are within the experimental error bounds \citep{lakebrink2018numerical,burrows2020evolution} for all simulated conditions. However, as shown in Figure \ref{fig:L7withmach}(c), the maximum (temporal) azimuthal distortion is underpredicted at all Mach numbers (consistently, Figure \ref{fig:maxdpcpmachall} shows that the predicted recovery patterns at the maximum distortion instances are qualitatively markedly different).  The predicted 
maximum distortion is likely affected 
due to the fact that the simulated time 
is two orders of magnitude shorter than the experimental time sequence.  For completeness, we also present the predicted maximum distortions from unfiltered, raw time sample from the wall-modeled LES. The maximum distortions increase significantly, however, are still underpredicted. In Appendix I, a brief application of the Extreme Value Theory \citep{migliorini2024evaluation,gil2018assessment,tanguy2018characteristics} to predict the extreme events of the probability distribution function of the azimuthal distortions is presented, with improved comparisons between the LES and the experiments. Appendix II demonstrates that in the experiments, the perturbations in $max(\mathrm{dPcP})$ occur at a frequency $1-2$ Hz, which are not sampled in the current simulations. Thus, it is conjectured that the extent of the under-prediction of the maximum flow distortion can, in part, also be attributed to the limited simulation integration times.  \\

\subsection{Statistical comparison of dynamic flow distortions}

\noindent
In this section, we plot quantile-quantile (Q-Q) plots of the pressure recovery ($\mathrm{PR}$), and ring-averaged azimuthal distortion ($\mathrm{dPcP}_i$ where $i \in \{1,2,3,4,5\}$, and $i=1$ and $i=5$ are the innermost and outermost rings, respectively between the experiments \citep{burrows2020evolution} and the predictions on the $L7$ grid. For brevity, the forty probe $\mathrm{PR}$ and the ring-averaged radial distortion ($\mathrm{dPrP}$,see definition in \#ARP1420-B SAE report) Q-Q plots are not presented in this article, but may be accessed in the supplementary material accompanying this article\footnote{Link to supplementary material: \href{https://ctr.stanford.edu/about-center-turbulence-research/research-data}{https://ctr.stanford.edu/about-center-turbulence-research/research-data}}.. \\

\noindent
A Q-Q plot resembles a identity line ($y=x$) if the experimental and simulation datasets are sampled from similar statistical distributions. If the two samples are otherwise linearly correlated, then, the Q-Q plot resembles a different straight line ($y \neq x$). For a Q-Q plot flatter than the identity line, the ``x-axis'' data has wider tails than the ``y-axis'' data and vice versa. Finally, an ``S'' shaped distribution indicates that one of the samples is more skewed than the other, or that one of the samples has heavier tails than the other. \\

\begin{figure}[!ht] 
\centering{
{\includegraphics[width=0.3\textwidth]{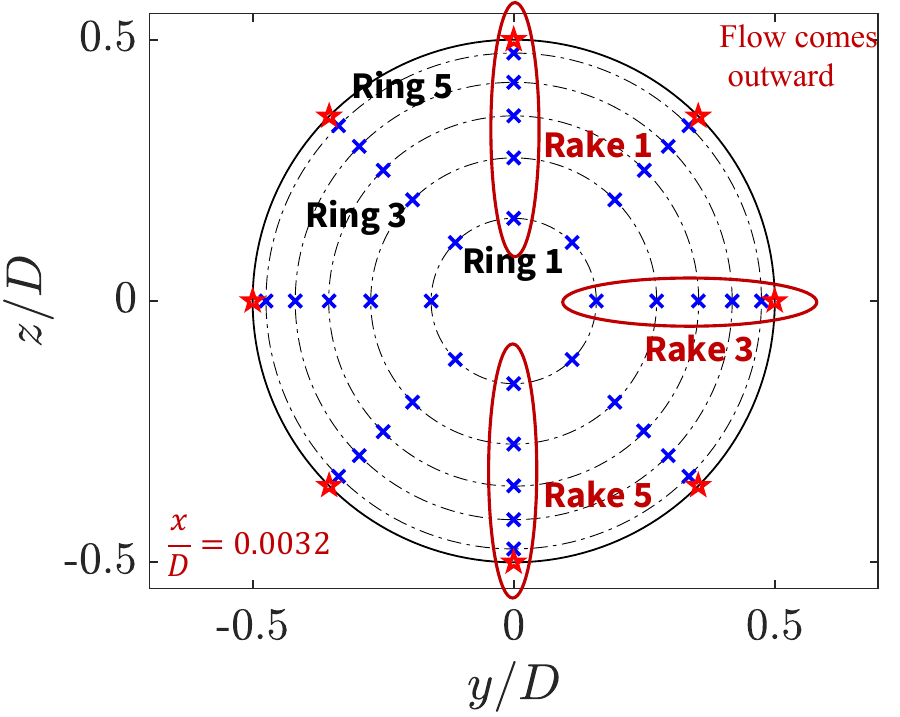}}} \\
(a) 
{\includegraphics[width=1.0\textwidth]{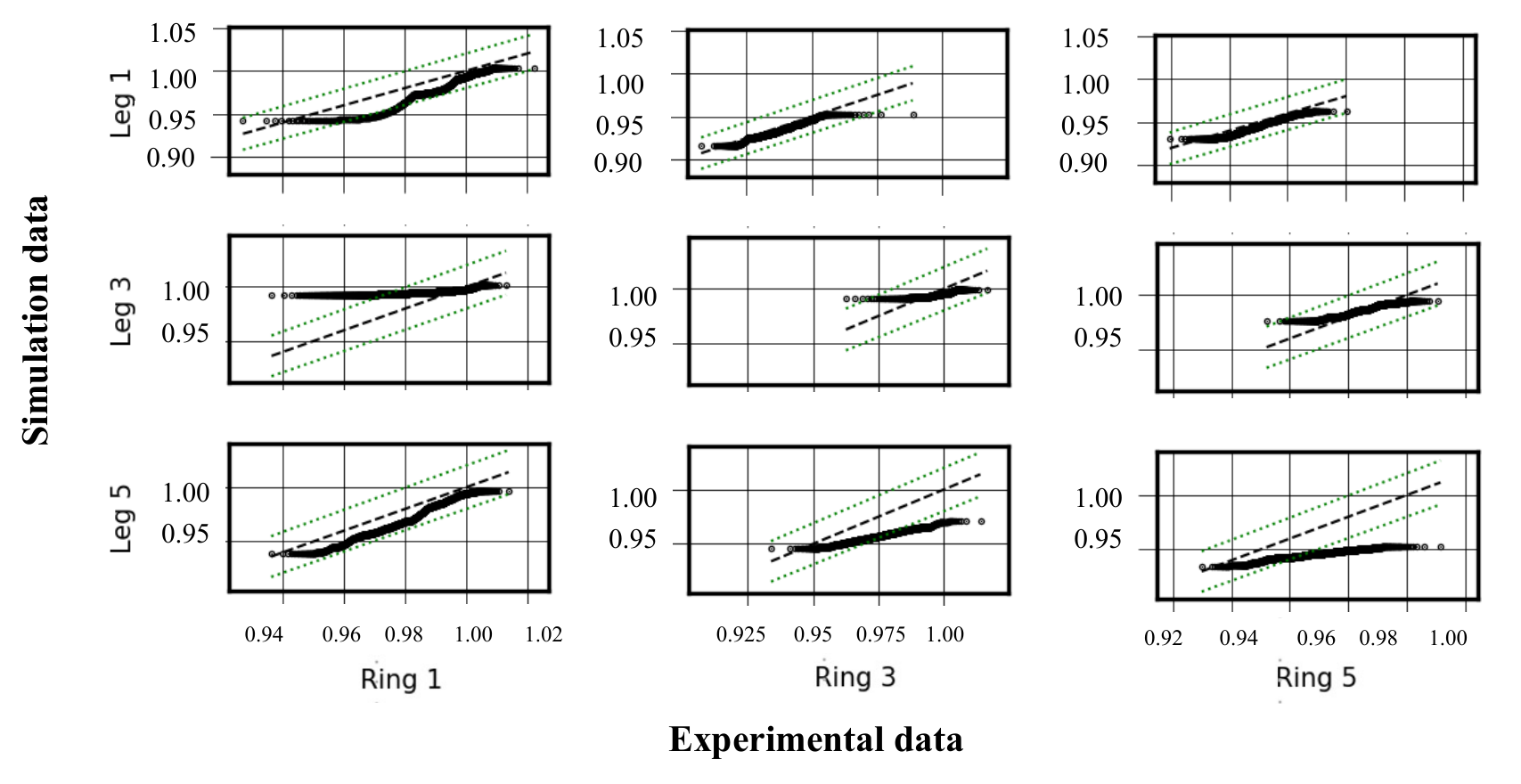}}    
(b)
\caption{ (a) A schematic of the positioning of the concentric rings {1, 3 and 5} and rakes/legs (on each ring) numbered 1, 3, and 5 (representative of the top wall separation, the core flow region, and the bottom wall separation, respectively). (b) Quantile-Quantile (Q-Q) plots (solid black lines) of pressure recovery, $\mathrm{PR}$, between LES (on the $L7$ grid with DTCSM/SWM models) and the experiments for the $Ma_{AIP} \sim 0.36$ condition. The dashed black line is the identity line ($y=x$) and the green dashed-dotted lines are the 2\% acceptance intervals per the SAE report \#AIR6345.}
     \label{fig:pr_qq_36}
  \end{figure}

\noindent
Figures \ref{fig:pr_qq_36} \textemdash \ref{fig:pr_qq_54} present the Q-Q plots of pressure recovery at the AIP with tolerance bounds of 2\% per the SAE 
standard (report \#AIR6345). In particular, three rakes/legs are considered, rakes/legs 1, 3 and 5 which correspond to the top center, middle, and the bottom center of the AIP plane (see Figure \ref{fig:pr_qq_36}(a)). For all Mach numbers considered, the Q-Q plots are generally within the accuracy tolerance bounds on the top wall suggesting that the dynamics of the top wall separation are well characterized in the simulation. In the core region, especially on Rings 1 and 3, the simulations predict a nearly constant pressure recovery, consistent with an overly inhibited dynamical distortions in the core flow. This may be partly attributed to the coarser grids in the core which dissipates some of the flow unsteadiness or weaker separations on the walls that do not lift up enough to reach and perturb the core. On the bottom wall, the dynamics of the pressure recovery on the innermost ring are well captured. However, on the outer-rings, the simulations under-sample the higher recovery events. Physically, the over and under sampling of the pressure recovery on rakes/legs 3 and 5, respectively, is also consistent with a reduced lift up of the vortices on the bottom wall toward the core.      

 \begin{figure}[!ht] 
\centering
{\includegraphics[width=0.95\textwidth]{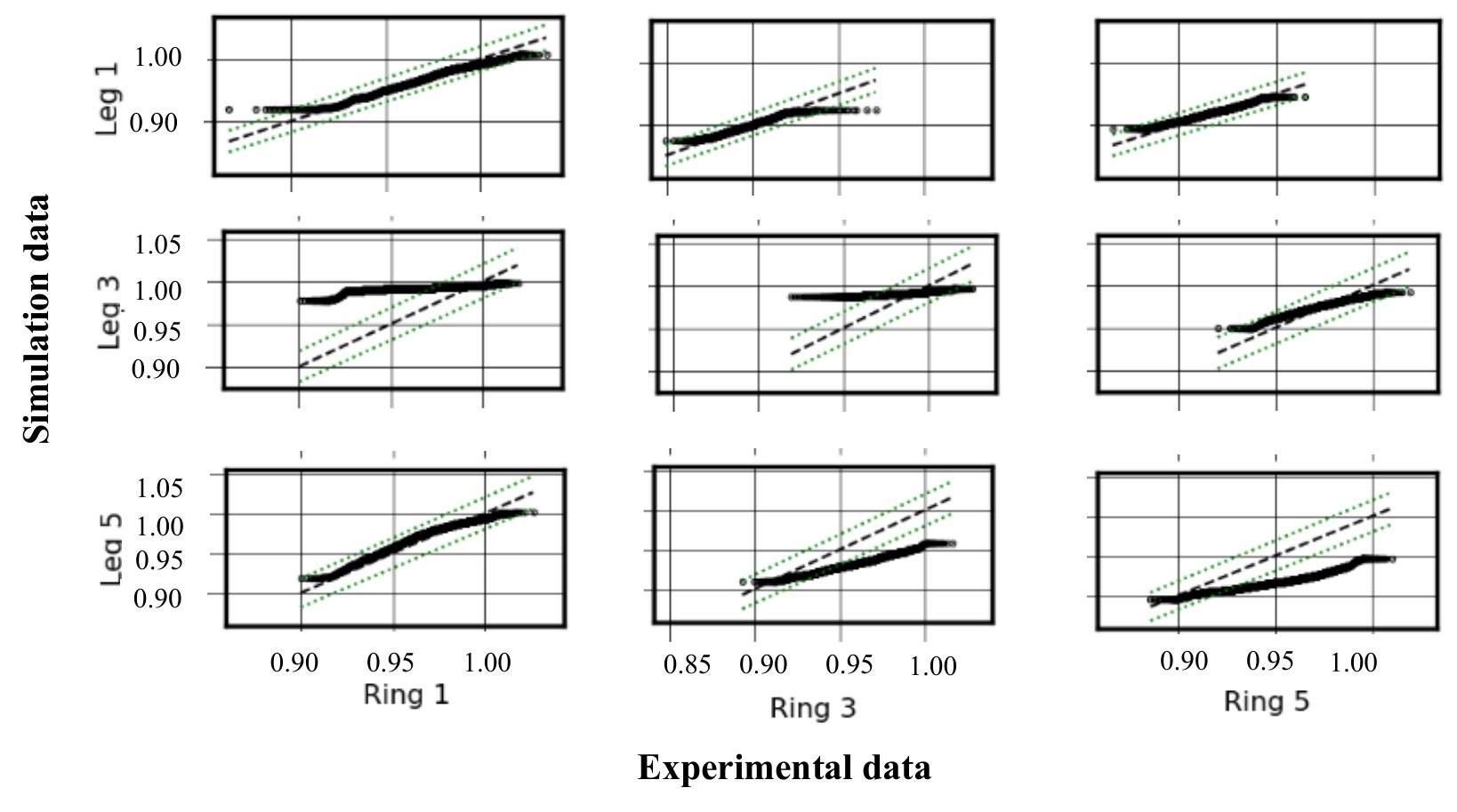}}    \caption{ Quantile-Quantile (Q-Q) plots (solid black lines) of pressure recovery, $\mathrm{PR}$, between LES (on the $L7$ grid with DTCSM/SWM models) and the experiments for the $Ma_{AIP} \sim 0.46$ condition. The dashed black line is the identity line ($y=x$) and the green dashed-dotted lines are the 2\% acceptance intervals per the SAE report \#AIR6345.}
     \label{fig:pr_qq_46}
  \end{figure}

\begin{figure}[!ht] 
\centering
{\includegraphics[width=0.95\textwidth]{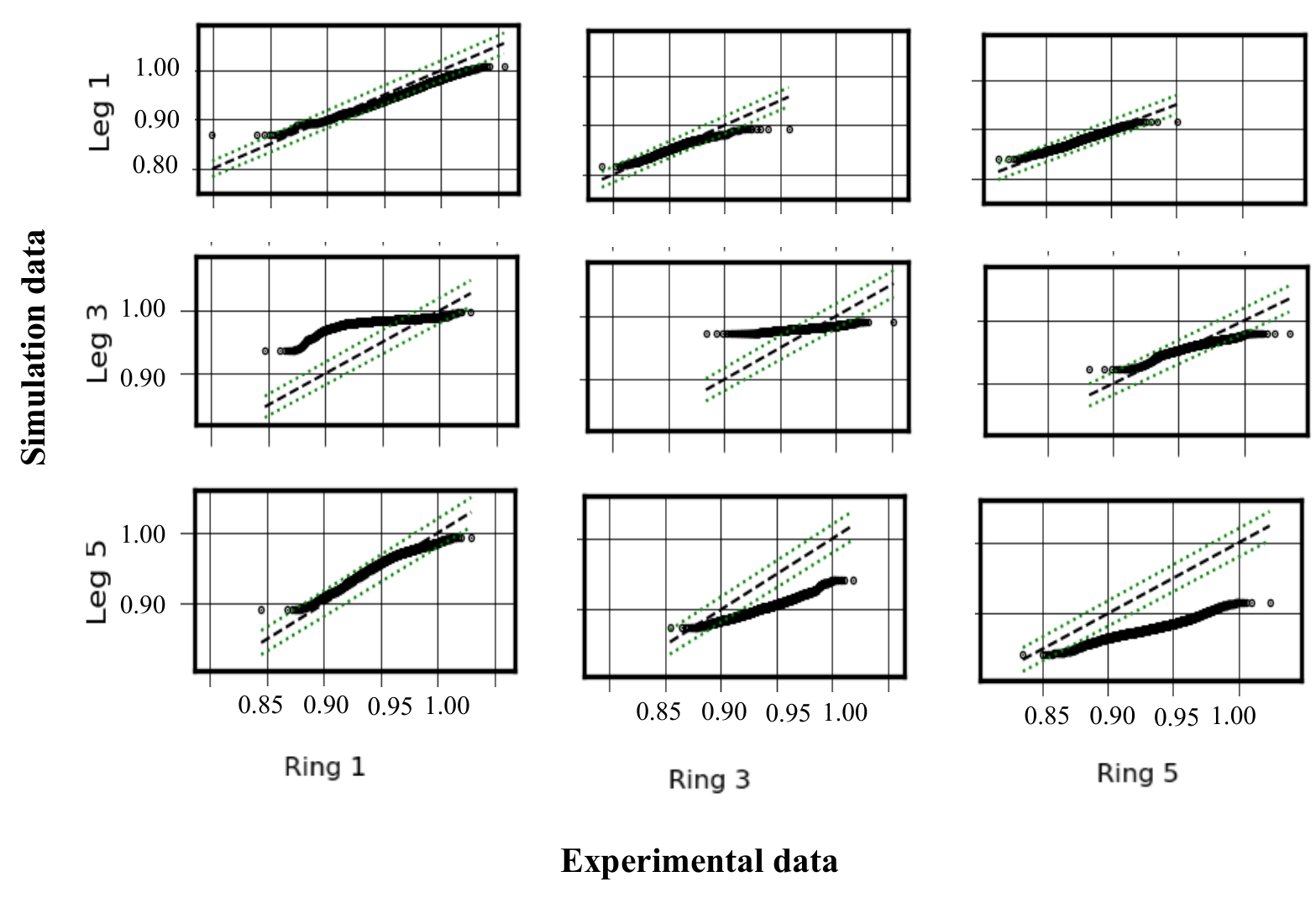}} 
    \caption{Quantile-Quantile (Q-Q) plots (solid black lines) of pressure recovery, $\mathrm{PR}$, between LES (on the $L7$ grid with DTCSM/SWM models) and the experiments for the $Ma_{AIP} \sim 0.54$ condition. The dashed black line is the identity line ($y=x$) and the green dashed-dotted lines are the 2\% acceptance intervals per the SAE report \#AIR6345.}
     \label{fig:pr_qq_54}
  \end{figure}

\begin{figure}[!ht] 
\centering
    {\includegraphics[width=1\textwidth]{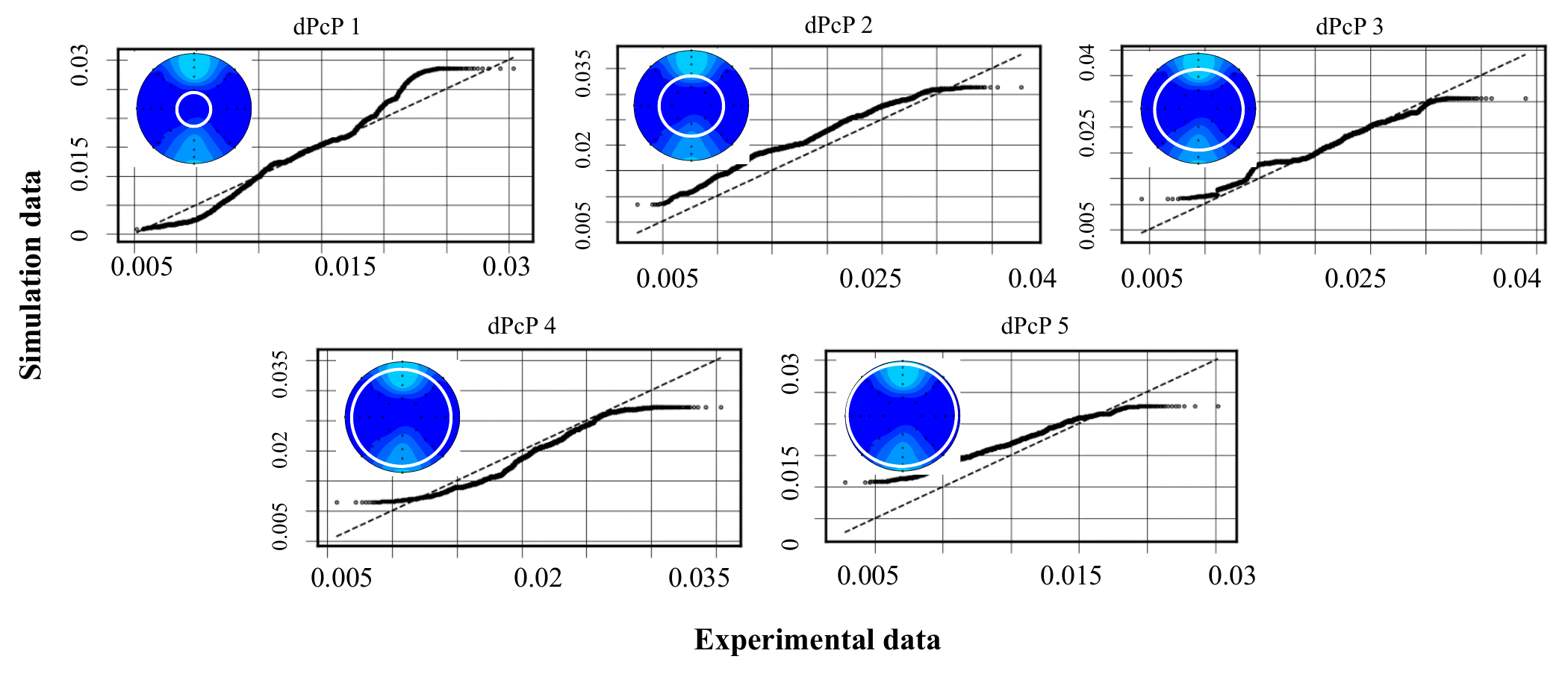}} 
    \caption{ Quantile-Quantile (Q-Q) plots (solid black lines) of $\mathrm{dPcP}$ between LES (on the $L7$ grid with DTCSM/SWM models) and the experiments of \citet{burrows2020evolution} for the $Ma_{AIP} \sim 0.36$ condition. The dashed black line is the identity line ($y=x$). } 
     \label{fig:ring3p55}
 \end{figure}

\noindent
Figures \ref{fig:ring3p55} \textemdash \ref{fig:ring4p85} show the $\mathrm{dPcP}_i$ Q-Q plots for the lowest, intermediate, and highest Mach number conditions, respectively. The solid black lines denote the data, and the dashed black line marks the reference identity line. 
For the lowest Mach number flow, the comparisons are generally favorable in the middle of the Q-Q plots, indicating that the experiments and the simulations have similar ring-averaged distortion values, and similar statistical distributions around the mean value. The left and right tails of the Q-Q plots are significantly flatter, suggesting that the simulation data samples more frequent low distortion and less frequent higher distortions potentially due to the smaller simulation integration time. For the outermost ring, the LES data suffers from a shallower distribution even around the mean. Overall, the Q-Q plots are generally within the 2\% tolerance bound around the identity line, except at the tails and the outermost ring (closest to the wall), where the separation is slightly under-predicted (see Figure \ref{fig:sd2dtcsmswmcp_3p55}). \\

\noindent
At the higher Mach numbers, notably the ring-averaged $\mathrm{dPcP}_i$ reaches larger instantaneous values due to the increased unsteadiness from the larger separations and weak shocks. For both $Ma_{AIP} \sim 0.46, \, 0.54$ cases, the Q-Q plot is generally close to the identity line, linearly correlating with experiments, also suggesting similar statistical distribution around the mean value. However, for the innermost ring, the simulations underpredict the average distortion. On the other rings, the simulations underpredict the higher distortion events, while overpredicting the frequency of smaller distortion events \footnote{Although not shown, visualizing the temporal history of the total pressure contour at the AIP reaffirms that the simulations underpredict the frequency of high distortion events, such as, instantaneous merging of the pressure losses at the bottom and top walls; which also affects the averaged distortion on the innermost ring.}.  \\

 \begin{figure}[!ht] 
\centering
    {\includegraphics[width=1\textwidth]{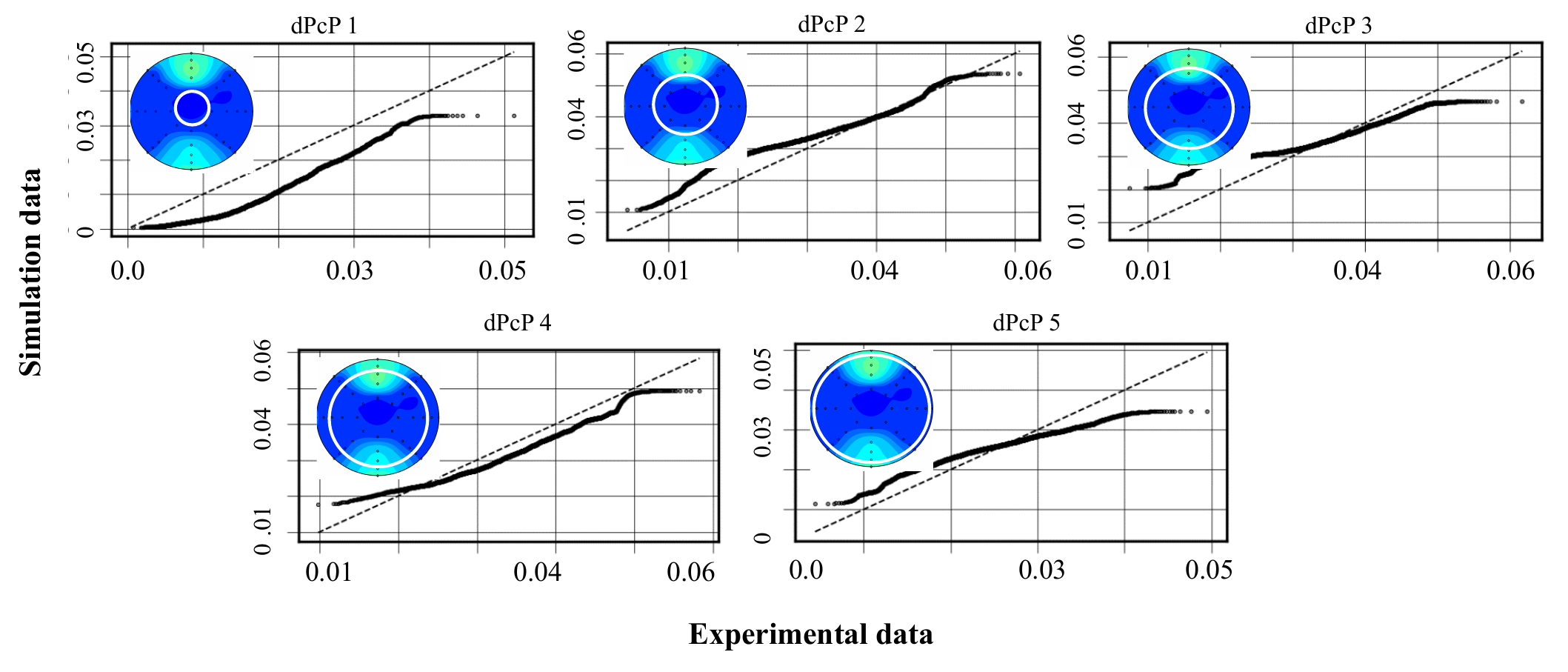}} 
    \caption{ Quantile-Quantile (Q-Q) plots (solid black lines) of $\mathrm{dPcP}$ between LES (on the $L7$ grid with DTCSM/SWM models) and the experiments of \citet{burrows2020evolution} for the $Ma_{AIP} \sim 0.46$ condition. The dashed black line is the identity line ($y=x$). } 
     \label{fig:ring4p31}
 \end{figure}

 \begin{figure}[!ht] 
\centering
    {\includegraphics[width=1\textwidth]{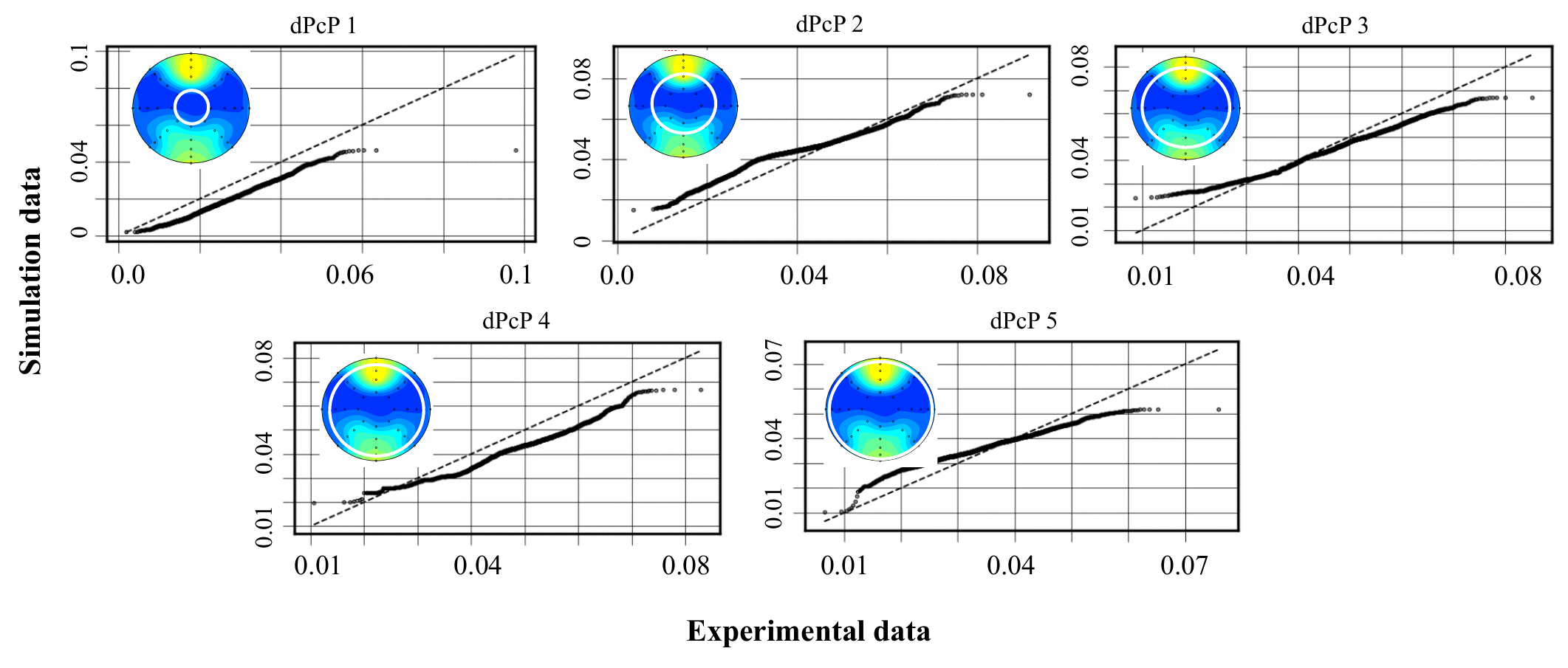}} 
    \caption{ Quantile-Quantile (Q-Q) plots (solid black lines) of $\mathrm{dPcP}$ between  LES (on the $L7$ grid with DTCSM/SWM models) and the experiments of \citet{burrows2020evolution} for the $Ma_{AIP} \sim 0.54$ condition. The dashed black line is the identity line ($y=x$). } 
     \label{fig:ring4p85}
 \end{figure}

\section{Concluding remarks}
\label{sec:conclude}
\noindent
This work examines the flow separation and the resulting pressure distortions inside the SD-2 serpentine diffuser from Georgia Tech \citep{burrows2020evolution} at three exit plane (AIP) Mach numbers, $Ma_{AIP} \approx \{ 0.36, \, 0.46, \, 0.54 \}$. As the Mach number increases, the incoming flow experiences a shock leading to a separation on the bottom wall. Slightly downstream, the flow undergoes a second separation on the top wall due to prolonged pressure gradients. \\  



\noindent
Wall-modeled LES are performed for all three Mach number flows, with a particular focus on the transonic condition.  The recently proposed dynamic tensor-coefficient Smagorinsky subgrid-scale \citep{agrawal2022non} and sensor-aided non-equilibrium wall models \citep{agrawal2024non} show improved predictions compared to the use of the dynamic Smagorinsky \citep{moin1991dynamic} and equilibrium wall closures \citep{cabot2000approximate}. With the newer models, the obtained pressure recovery and azimuthal flow distortion at the exit plane are in good agreement with the experiments (to within $0.3\%$ and 2.5\% for averaged recovery and distortions, respectively, which are both within the experimental error bounds).  \\

\noindent
At the two lower Mach numbers, $Ma_{AIP} \approx \{0.36, \, 0.46\}$, simulations with DTCSM/SWM reasonably predict the surface pressure, $C_p$, on the $L7$ grid.
The resolution provided by the $L7$ grid
is consistent with theoretical estimates for requisite resolution for wall modeled LES for separating flows \cite{agrawal2023grid} ($\Delta/min(l_p) \lesssim \mathcal{O}(10)$). The time-averaged pressure recovery contours at the AIP show increasing separations on the top and the bottom walls with increasing Mach number. At the instance of maximum azimuthal distortion, the separation patterns on the two walls are significantly larger than the mean, and engulf the core region, especially at the highest Mach number. The time-averaged pressure recovery and azimuthal distortion values compare well with the experiments for all Mach numbers (to within one standard deviation). However, the maximum distortion values are underpredicted. We conjecture that this may partly be attributed to the much longer (and hence not-simulated) time-scales over which the maximum distortion changes in the experiments. \\

\noindent
A statistical analysis of the ring-averaged azimuthal distortions via the quantile-quantile plots is performed. For the lowest Mach number flow, the comparisons are generally favorable in the middle of the Q-Q plots; but the left and right tails are significantly flatter. These suggest that the simulations resembles the statistical distribution of the experiments around the mean azimuthal distortion, but samples more frequent low distortion and less frequent higher distortions (potentially due to the smaller integration time, and/or due to diminished unsteady perturbations around the mean state in the simulations). At the higher Mach numbers, the ring-averaged $\mathrm{dPcP}_i$ are instantaneously larger. The Q-Q plot resembles a linear correlation, suggesting a similar statistical distribution around the mean. However, for the innermost ring, the simulations slightly underpredict the average distortion. \\



\section*{Appendix I: Application of Extreme Value Theory for predictions of maximum distortion}
\label{sec:evt}
\vspace{5pt}
\noindent
\citet{migliorini2024evaluation,gil2018assessment,tanguy2018characteristics} have previously utilized the (statistical) extreme value theory (EVT) to provide improved predictions of the maximum flow distortions, since the simulation integration times are $\mathcal{O}(10-100)$ smaller than experiments. Herein, the EVT assumes a distribution form of the tails of the sampled LES $\mathrm{dPcP}$ values. We utilize the standard, \textit{ MATLAB evfit} functionality with an (arbitrarily) chosen upper probability threshold of 0.1\% to model the value of the maximum azimuthal distortion.  \\

\begin{figure}[!ht] 
\centering
    {\includegraphics[width=0.4\textwidth]{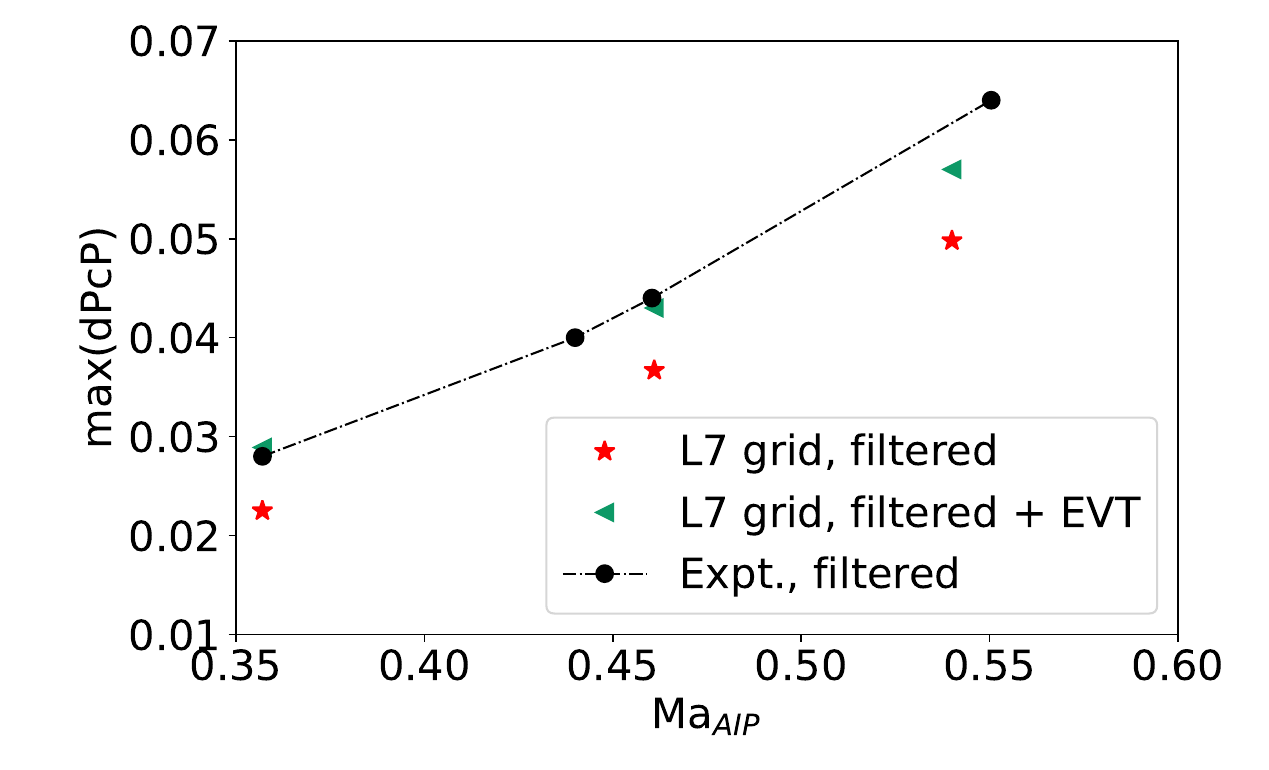}} 
    \caption{ Predictions of the maximum value of face averaged flow distortion with Mach number for the flow inside the SD-2 diffuser, including those using Extreme Value theory.  }
     \label{fig:evt}
 \end{figure}

\noindent
The functional form of the assumed probability distribution function used in \textit{evfit} resembles a Gumbel distribution, and is given as, 

\begin{equation}
    y = f(x \vert \mu, \, \sigma) = \frac{1}{\sigma} exp \bigg[ \frac{(x-\mu)}{\sigma}) \bigg] exp \bigg[ - exp \big[ \frac{(x-\mu)}{\sigma}) \big] \bigg] 
\end{equation}
where $\mu, \, \sigma$ are the parameter estimates from the sampled LES data. Based on this approach, Figure \ref{fig:evt} demonstrates that EVT application increases the predicted $\mathrm{dPcP}$. For the two lower Mach numbers, the predicted values are in agreement with the experiment, but for the transonic throat flow, the EVT applied distortion value is lower than the experiment. The remaining mispredictions on the highest Mach number flow may highlight an under-prediction of the large distortion events (in the LES) to the extent that the EVT cannot sample relevant intermittent events to provide accurate maximum distortions. \\

\noindent
Finally, although not shown, we remark that the peak distortion from the EVT is dependent on the probability threshold (e.g.: the predicted peak $\mathrm{dPcP}$ decreases by 12\% for the highest Mach number flow if the threshold is increased to 0.5\%).

 \section*{Appendix II: Running time average of the maximum azimuthal distortion in experiments}
\label{sec:runtimeavgexpt}
\vspace{5pt}

\begin{figure}[!ht] 
\centering
(a){\includegraphics[width=0.45\textwidth]{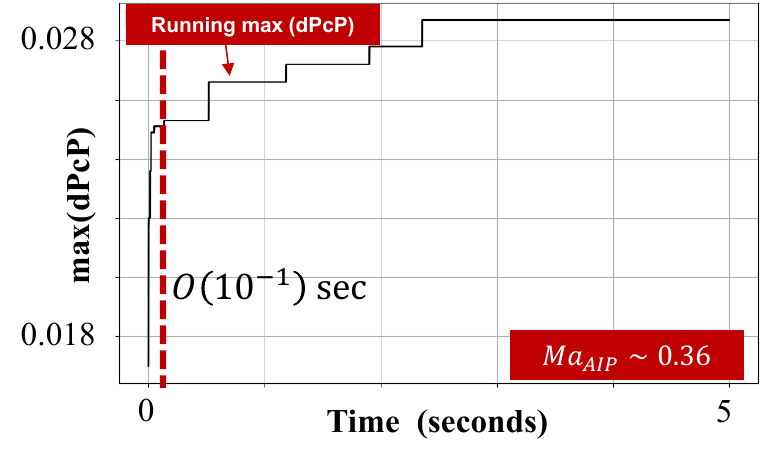}} 
(b){\includegraphics[width=0.45\textwidth]{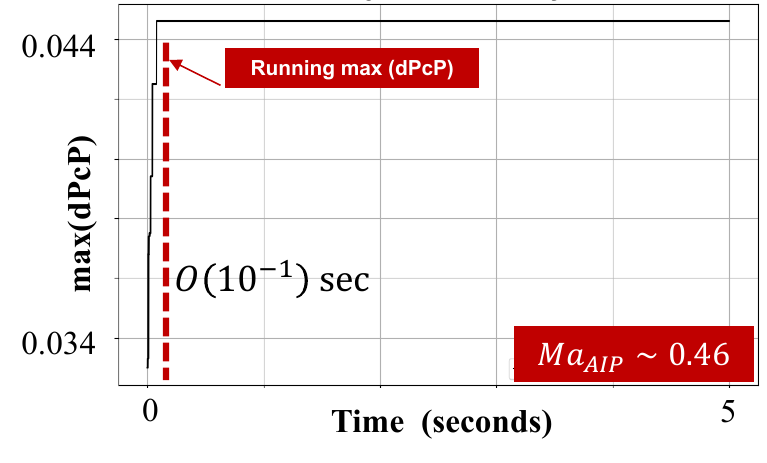}} \\
(c){\includegraphics[width=0.5\textwidth]{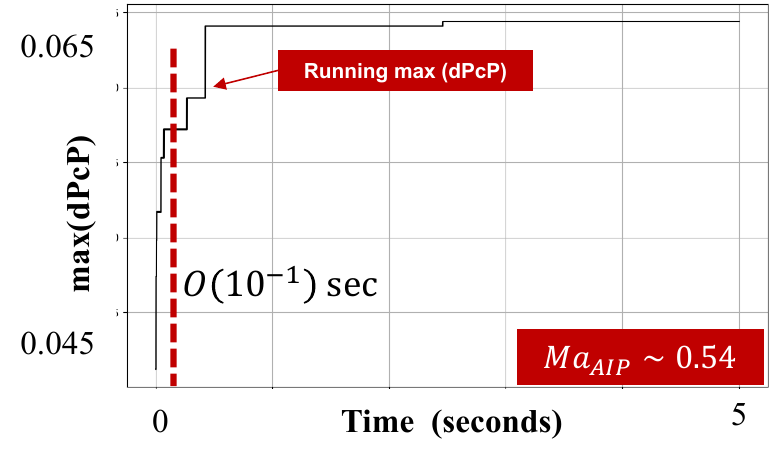}} 
    \caption{ The running time averages of the maximum azimuthal distortion, $max(\mathrm{dPcP})$ in the experiments for the flow inside the SD-2 diffuser at (a) $Ma_{AIP} \sim 0.36$, (b) $Ma_{AIP} \sim 0.46$ and (c) $Ma_{AIP} \sim 0.54$ conditions respectively.  }
     \label{fig:runtime}
 \end{figure}

\noindent
As the simulation time horizon is much smaller than the experiments, in this appendix, we briefly report the moving time average of the maximum distortion observed in the experiments. Figure \ref{fig:runtime} demonstrates that changes in the running maximum value of $\mathrm{dPcP}$ occur at a frequency of $\mathcal{O}(1)$ Hz (which are not sampled in the current simulations. The simulations were computed for smaller integration times than $\mathcal{O}(10^{-1})$ sec).

\section*{Acknowledgments}
\noindent
This work was supported by Boeing Research and Technology under grant \#2024-UI-PA-100. R.A. also acknowledges support from the Franklin P. and Caroline M. Johnson Graduate Fellowship at Stanford University. This research used resources of the Oak Ridge Leadership Computing Facility, which is a DOE Office of Science User Facility supported under Contract DE-AC05-00OR22725. We also acknowledge fruitful discussions with Michael Whitmore at Stanford University.

\end{document}